\documentclass[onecolumn,showpacs,preprintnumbers,amsmath,amssymb,aps,prb,superscriptaddress]{revtex4-2}

\pdfoutput=1
\usepackage[pdftex]{graphicx}
\usepackage[english]{babel}
\usepackage{soul}
\usepackage{cleveref}
\usepackage{bbold}
\usepackage{mathtools}
\usepackage{multirow}
\usepackage{colortbl}
\definecolor{kugray5}{RGB}{224,224,224}
\usepackage[latin1]{inputenc}
\usepackage{diagbox}
\usepackage{mciteplus}
\usepackage{mathrsfs}
\usepackage{amsmath}

\usepackage{dcolumn}% Align table columns on decimal point\right] 
\usepackage{bm}% bold math

\usepackage{color}
\usepackage{amstext}
\usepackage{braket}
\usepackage{mathdots}
\usepackage{tikz}
\usepackage{physics}
\usepackage{enumitem}
\usepackage{siunitx}

\usetikzlibrary{matrix,decorations.pathreplacing}

\begin{document}

%%%%%%%%%%%%%%%%%%%%%%%%%%%%%%%%%%%%%%%%%%%%%%%%%%%%%%%%%%%%%%%%%%%%%%%%%%
%                               Title                                    %
%%%%%%%%%%%%%%%%%%%%%%%%%%%%%%%%%%%%%%%%%%%%%%%%%%%%%%%%%%%%%%%%%%%%%%%%%%

\title{Topological $n$-root Su-Schrieffer-Heeger model in a non-Hermitian photonic ring system}

\author{David Viedma}
\email{david.viedma@uab.cat}
\affiliation{Departament de F\'isica, Universitat Aut\`onoma de Barcelona, E-08193 Bellaterra, Spain}

\author{Anselmo M. Marques}
\affiliation{Department of Physics $\&$ i3N, University of Aveiro, 3810-193 Aveiro, Portugal}

\author{Ricardo G. Dias}
\affiliation{Department of Physics $\&$ i3N, University of Aveiro, 3810-193 Aveiro, Portugal}

\author{Ver\`onica Ahufinger}
\affiliation{Departament de F\'isica, Universitat Aut\`onoma de Barcelona, E-08193 Bellaterra, Spain}

%\date{\today}

%%%%%%%%%%%%%%%%%%%%%%%%%%%%%%%%%%%%%%%%%%%%%%%%%%%%%%%%%%%%%%%%%%%%%%%%%%
%                              abstract                                  %
%%%%%%%%%%%%%%%%%%%%%%%%%%%%%%%%%%%%%%%%%%%%%%%%%%%%%%%%%%%%%%%%%%%%%%%%%%

\begin{abstract}
Square-root topology is one of the newest additions to the ever expanding field of topological insulators (TIs).
It characterizes systems that relate to their parent TI through the squaring of their Hamiltonians.
Extensions to $2^n$-root topology, where $n$ is the number of squaring operations involved in retrieving the parent TI, were quick to follow.
Here, we go one step further and develop the framework for designing general $n$-root TIs, with $n$ any positive integer, using the Su-Schrieffer-Heeger (SSH) model as the parent TI from which the higher-root versions are constructed.
The method relies on using loops of unidirectional couplings as building blocks, such that the resulting model is non-Hermitian and embedded with a generalized chiral symmetry.
Edge states are observed at the $n$ branches of the complex energy spectrum, appearing within what we designate as a ring gap, shown to be irreducible to the usual point or line gaps.
We further detail on how such an $n$-root model can be realistically implemented in photonic ring systems.
Near perfect unidirectional effective couplings between the main rings can be generated via mediating auxiliary rings with modulated gains and losses.
These induce high imaginary gauge fields that strongly supress couplings in one direction, while enhancing them in the other.
We use these photonic lattices to validate and benchmark the analytical predictions.
Our results introduce a new class of high-root topological models, as well as a route for their experimental realization.
\end{abstract}

%\pacs{74.25.Dw,74.25.Bt}

\maketitle

% GUIDELINES
%%%%%%%%%%%%%%%%%%%%%%%%%%%%%%%%%%%%%%%%%%%%%%%%%%%%%%%%%%%%%%%%%%%%%
%%%%%%%%%%%%%%%%%%%%%%%%%%%%%%%%%%%%%%%%%%%%%%%%%%%%%%%%%%%%%%%%%%%%%
%%%%%%%%%%%%%%%%%%%%%%%%%%%%%%%%%%%%%%%%%%%%%%%%%%%%%%%%%%%%%%%%%%%%%
% \section{Guidelines for LSA}

% Sections: Introduction, Results, Discussion, Materials and Methods, Acknowledgements etc

% Abstract: Max 250 words

% Body: Max 6000 words (not counting figs, etc)

% References: Max 50

% MS WORD

% https://www.nature.com/documents/lsa-gta.pdf

% SECTION
%%%%%%%%%%%%%%%%%%%%%%%%%%%%%%%%%%%%%%%%%%%%%%%%%%%%%%%%%%%%%%%%%%%%%
%%%%%%%%%%%%%%%%%%%%%%%%%%%%%%%%%%%%%%%%%%%%%%%%%%%%%%%%%%%%%%%%%%%%%
%%%%%%%%%%%%%%%%%%%%%%%%%%%%%%%%%%%%%%%%%%%%%%%%%%%%%%%%%%%%%%%%%%%%%
\section{Introduction}
\label{sec:intro}

High-root topology has emerged as a rich new branch within the field of topological insulators (TIs). 
Square-root TIs ($\sqrt{\text{TIs}}$) \cite{Arkinstall2017} were first proposed to characterize lattice models whose parent TI, from which its topological features are inherited, manifests itself as one of the diagonal blocks of the squared Hamiltonian \cite{Kremer2020,Pelegri2019,Ke2020,Yoshida2021,Ding2021}.
Experimental realization of these models in different platforms followed soon \cite{Yan2020,Song2020,Yan2021,Song2022,Cheng2022,Kang2023,Wu2023,Yan2023,Guo2023}.
Subsequently, these systems were further generalized to $2^n$-root TIs ($\sqrt[2^n]{\text{TIs}}$) \cite{Dias2021,Marques2021,Marques2021b,Marques2023}, meaning models that connect to their parent TI through a sequence of $n$ squaring operations.
The first experimental demonstrations of quartic-root topology ($n=2$) appeared recently in the context of acoustic \cite{Cui2023} and photonic \cite{Wei2023} lattices.
Studies on related topics, such as those of fractionally twisted models \cite{Basa2022} or multiplicative topological phases \cite{Cook2022}, have also started to appear recently.

The question of whether general $n$-root TIs ($\sqrt[n]{\text{TIs}}$), with $n\in\mathbb{N}$, can be devised naturally arises.
It has already been affirmatively answered for Floquet systems \cite{Bomantara2022,Zhou2022}, through a method based on subdividing the driving period into $n$ subperiods, each with its own associated Hamiltonian.
However, for non-driven systems, a natural generalization of $\sqrt[2^n]{\text{TIs}}$ to $\sqrt[n]{\text{TIs}}$ has been lacking so far.
Here, we bridge this gap in the literature by considering the SSH model \cite{Su1979} as the parent TI, from which its higher-root versions ($\sqrt[n]{\text{SSH}}$) are derived following a novel procedure.
Specifically, it involves constructing $n$-partite chains from loop modules of unidirectional couplings as the building blocks.
Under open boundary conditions (OBC), $n$ edge states, all decaying from the same edge \cite{Marques2021}, are seen to appear in the complex energy spectrum when in the topological phase.

The main challenge regarding the experimental design of the $\sqrt[n]{\text{SSH}}$ model relates to the implementation of the unidirectional couplings in the loop modules. Although seemingly exotic, non-Hermitian couplings have been a matter of intense discussion in recent years, with theoretical proposals and experimental implementations appearing in many different platforms, including optical and acoustic ring resonators \cite{Longhi2015,Lin2021,Lin2021OL,Zhang2021,Gu2022,Gao2022,Liu2022,Gao2023}, optical fibers and waveguides \cite{Weidemann2020,Weidemann2022,Ke2023}, ultracold atoms \cite{Gong2018,Liu2019}, electrical circuits \cite{Hofmann2019,Helbig2020,Liu2021,Zou2021}, modulated waveguides exploiting synthetic dimensions \cite{Qin2020,Zheng2022}, and many others.

%\dav{Using rings for real flux (\cite{Hafezi2011,Hafezi2013,Mittal2014}...) and imaginary flux (\cite{Longhi2015,Lin2021,Lin2021_PRA,Zhang2021}...). } %Mittal2006 https://www.nature.com/articles/nphoton.2016.10 for flux via heater or electro-optic modulator
%Non-H system in waveguides Ke2023 PRA 107, 053508, Loss-induced nonreciprocity Huang2021 Light: Science & Applications 10, 30

We focus here on photonic ring systems, and show that they are a well suited candidate for the realization of these models. We consider an array made up of a set of resonant optical ring resonators, which constitute the main rings of the lattice, coupled through smaller antiresonant link rings, as illustrated in Fig.~\ref{fig:1}. 
The link rings feature a split gain/loss distribution, in which the upper half of the ring has gain characterized by a parameter $h$ while the lower half has an equal amount of loss. 
To avoid reflection effects, we use a sine-like distribution for the imaginary part of the refractive index. The antiresonant condition for a ring mode with propagation constant $\beta$ reads: $\beta \left(L_L-L_M\right) = (2m+1)\pi$, where $L_M$ and $L_L$ are the lengths of main and link rings, respectively, and $m$ is the circulation. 
Here, we will restrict ourselves to the clockwise ($m=1$) and counter-clockwise ($m=-1$) circulations. 
Through the presence of the link rings, and due to their balanced gain and loss distribution, an effective asymmetric coupling is enabled between the same circulation $m$ in the main rings $t_\pm = t \, e^{\pm h}$ \cite{Longhi2015}, which depends exponentially on the gain and loss parameters and is analogous to an imaginary gauge field acting on the system. 
We represent the forward coupling direction by $+$ and the backward direction by $-$. 
Unidirectionality in the couplings is obtained in the limit $h\to\infty$, while Hermiticity is restored for $h=0$.
For a finite $h$ value, one is in the intermediate situation where the hoppings occur in both directions, but with the predominance of one over the other.
For a strong enough gauge field, nearly perfect unidirectionality can be achieved, as we propose below. 
The coupling $t$ is determined by the relative distance between the main rings, which for the roots of the SSH model alternates between two values in different plaquettes to achieve $\sqrt[3]{t_1}\neq \sqrt[3]{t_2}$ (this choice of hopping notation \cite{Marques2021,Marques2021b,Cui2023}, adopted from now on, will become clear below, when we relate the model in Fig.~\ref{fig:1} to its cubed SSH parent system). 
A key characteristic in this system is that for each pair of rings, only opposite circulations may be coupled between them. That is, we consider that the coupling of the (counter-)clockwise circulation of a main ring with the (counter-)clockwise of a link ring is negligible. In that sense, we can separate the system in clockwise and counter-clockwise components for all main rings. 
This assumption is valid as long as the coupling region between main and link rings is long compared to the wavelength of light \cite{Hafezi2011}, which is fulfilled for the sizes considered in this work and is reflected in the numerical results. Additionally, a real flux of desired value can be established in ring systems by orthogonally displacing a link ring from the line connecting the centers of the corresponding main rings, and thus generating a phase in the coupling between them \cite{Hafezi2013}.

%The rest of the paper is organized as follows.

% SECTION
%%%%%%%%%%%%%%%%%%%%%%%%%%%%%%%%%%%%%%%%%%%%%%%%%%%%%%%%%%%%%%%%%%%%%
%%%%%%%%%%%%%%%%%%%%%%%%%%%%%%%%%%%%%%%%%%%%%%%%%%%%%%%%%%%%%%%%%%%%%
%%%%%%%%%%%%%%%%%%%%%%%%%%%%%%%%%%%%%%%%%%%%%%%%%%%%%%%%%%%%%%%%%%%%%

\section{Results}
\label{sec:results}

We will begin this section with a brief overview of the $\sqrt[3]{\text{SSH}}$ model and its main features.
The interested reader is referred to Supplementary Section~\ref{suppl-sec:ssh3} for the complete analytical description of the model.
Next, we will detail on how such a model can be implemented in a photonic ring system.
After a brief discussion on the generalized $\sqrt[n]{\text{SSH}}$ model, we will end this section with an analysis of the photonic realization of the $\sqrt[4]{\text{SSH}}$ model.

\subsection{$\sqrt[3]{\text{SSH}}$ model}
\label{subsec:ssh3}

\begin{figure}[t]
	\begin{centering}
		\includegraphics[width=0.45\columnwidth]{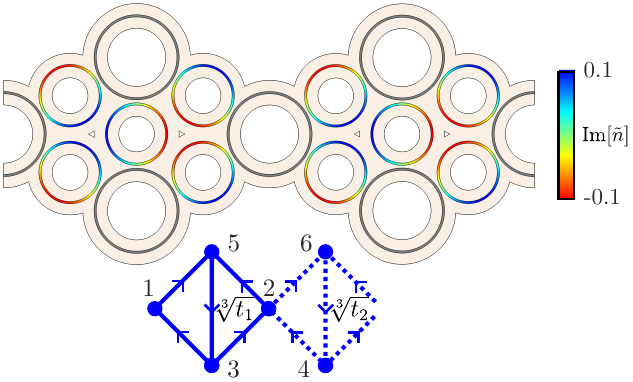} 
		\par\end{centering}
	\caption{Unit cell geometry of the photonic ring implementation of the $\sqrt[3]{\text{SSH}}$ model. The grey rings constitute the main rings of the effective lattice, without gain or loss. The smaller link rings are antiresonant to the former, and display a sine-like distribution of the imaginary component of the refractive index $\tilde{n}$, as represented by the color bar on the right. The distance between rings is different in each plaquette so that $\sqrt[3]{t_1}\neq\sqrt[3]{t_2}$. The lower inset depicts the unit cell of the $\sqrt[3]{\text{SSH}}$ model, where the arrows indicate the direction of the couplings, and corresponds to an effective description of the counter-clockwise ($m=-1$) circulation of the photonic system above. For the opposite clockwise ($m=1$) circulation, an equivalent model is obtained, but with all coupling directions flipped.}
	\label{fig:1}
\end{figure}

The unit cell of the $\sqrt[3]{\text{SSH}}$ model is depicted at the bottom of Fig.~\ref{fig:1}.
Under periodic boundary conditions (PBC), and in the ordered $\{\ket{j(k)}\}$ basis, with $j=1,2,\dots,6$, the bulk Hamiltonian of the $\sqrt[3]{\text{SSH}}$ model can be written as
\begin{eqnarray}
	H_{\sqrt[3]{\text{SSH}}}(k)&=&
	\begin{pmatrix}
		0&h_1&0
		\\
		0&0&h_2
		\\
		h_3&0&0
	\end{pmatrix},
	\label{eq:hamiltssh3}
	\\
	h_1&=&h_3^\dagger=-
	\begin{pmatrix}
		\sqrt[3]{t_1}&\sqrt[3]{t_2}e^{-ik}
		\\
		\sqrt[3]{t_1}&\sqrt[3]{t_2}
	\end{pmatrix},
	\label{eq:h1}
	\\
	h_2&=&-
	\begin{pmatrix}
		\sqrt[3]{t_1}&0
		\\
		0&\sqrt[3]{t_2}
	\end{pmatrix},
	\label{eq:h2}
\end{eqnarray}
where the lattice spacing was set to unity and all hopping terms are unidirectional.
We further assume $t_1,t_2\geq 0$, without loss of generality. 
Due to its tripartite nature, composed by the sublattices $(1,2)$, $(3,4)$ and $(5,6)$, and defined by requiring multiples of three hopping processes to produce intra-sublattice couplings \cite{Marques2022}, this Hamiltonian obeys a generalized chiral symmetry,
\begin{eqnarray}
	\mathscr{C}_3:\ \ &\Gamma_3&H_{\sqrt[3]{\text{SSH}}}(k)\Gamma_3^{-1}=\omega_3^{-1}H_{\sqrt[3]{\text{SSH}}}(k),
	\label{eq:genchiral}
	\\
	&\Gamma_3&=\text{diag}(\sigma_0,\omega_3\sigma_0,\omega_3^{-1}\sigma_0),
\end{eqnarray}
with $\omega_3=e^{i\frac{2\pi}{3}}$ and $\sigma_0$ the $2\times2$ identity matrix.

\begin{figure*}[t]
	\begin{centering}
		\includegraphics[width=0.975 \textwidth]{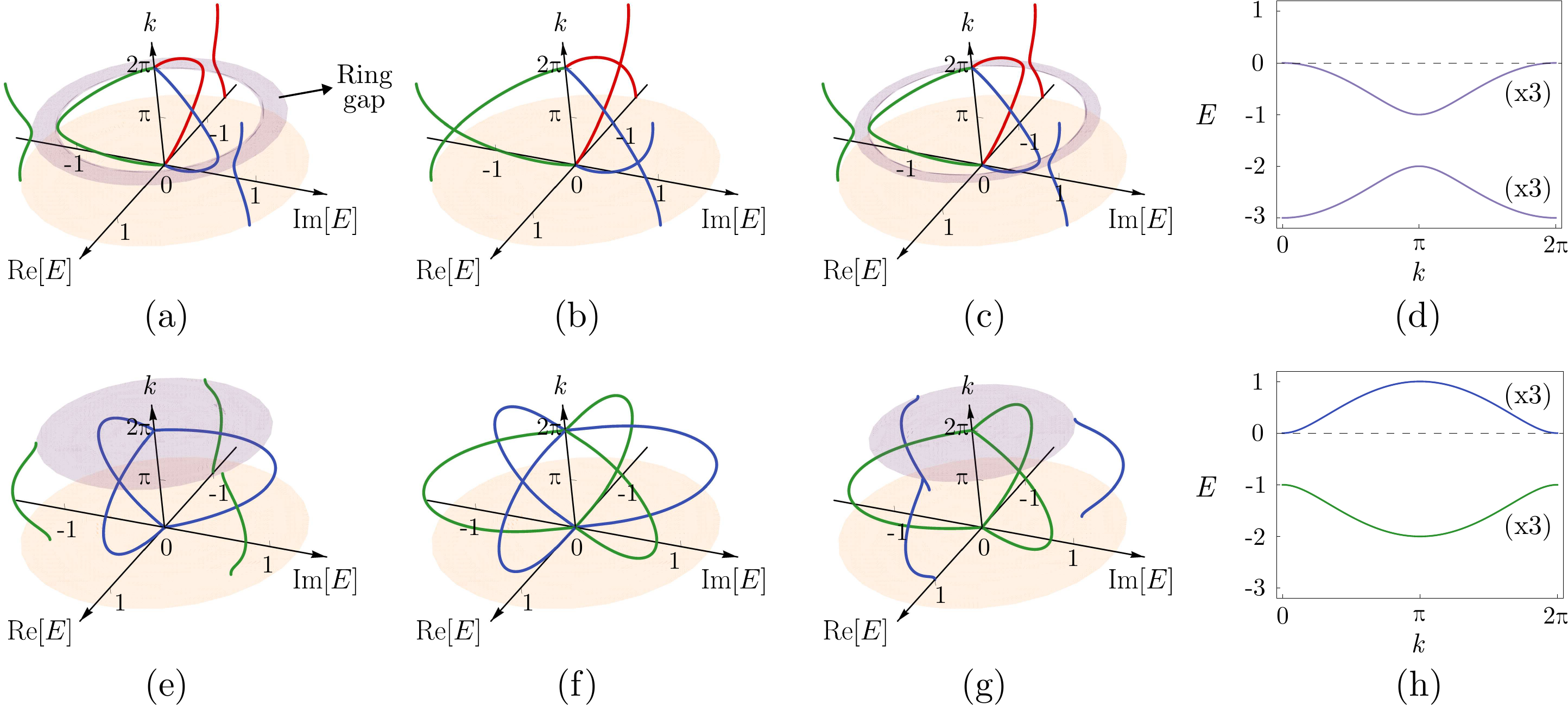} 
		\par\end{centering}
	\caption{Complex energy spectrum, in units of $\sqrt[3]{t_2}=1$, as a function of the momentum for the $\sqrt[3]{\text{SSH}}$ model in (\ref{eq:hamiltssh3}) with (a) $\sqrt[3]{t_1}=\sqrt[3]{0.5}$, (b) $\sqrt[3]{t_1}=1$, and (c) $\sqrt[3]{t_1}=\sqrt[3]{1.5}$. Different energy branches are indicated with different colors. (d) Cubed energy spectrum of the model in (a), which is purely real and with each band three-fold degenerate. (e)-(h) Same as the corresponding cases above, but for $\sqrt[3]{t_1}\to e^{i\frac{\pi}{3}}\sqrt[3]{t_1}$ in (\ref{eq:hamiltssh3}), leading to the $\sqrt[3]{\text{SSH}}_{\frac{\pi}{3}}$ model. Different colors in (e)-(g) now distinguish the groups of three bands that become degenerate upon cubing the spectrum. The ring gaps are depicted in light purple and appear at $k=\pi$ in (a) and (c), and at $k=2\pi$ in (e) and (g), where the inner circumference of the ring reduces to a point at $E=0$.}
	\label{fig:2}
\end{figure*}

After cubing the Hamiltonian in (\ref{eq:hamiltssh3}) we obtain
\begin{equation}
	H_{\sqrt[3]{\text{SSH}}}^3(k)=\text{diag}\Big(H_{\text{SSH}^\prime}(k),H_2(k),H_3(k)\Big),
	\label{eq:hamiltssh33}
\end{equation}	
where
\begin{eqnarray}
	H_{\text{SSH}^\prime}(k)&=&h_1h_2h_3 \nonumber
	\\
	&=&-
	\begin{pmatrix}
		t_1+t_2&t_1+t_2e^{-ik}
		\\
		t_1+t_2e^{ik}&t_1+t_2
	\end{pmatrix}  \nonumber
	\\
	&=&-(t_1+t_2)\sigma_0+H_{\text{SSH}}(k),
\end{eqnarray}
is isospectral to the other diagonal terms in (\ref{eq:hamiltssh33}), namely $H_2(k)=h_2h_3h_1$ and $H_3(k)=h_3h_1h_2$ \cite{Marques2022}.
Their eigenvalues are given by
\begin{equation}
	E_{\pm}(k)=-t_1-t_2\pm\sqrt{t_1^2+t_2^2+2t_1t_2\cos k}.
	\label{eq:sshspectrum}
\end{equation}
The three-fold degenerate spectrum of Fig.~\ref{fig:2}(d) is a reflection of the isospectrality of the three diagonal blocks.

The complex energy spectrum of the $\sqrt[3]{\text{SSH}}$ model, with the Hamiltonian in (\ref{eq:hamiltssh3}), is composed of three two-band branches that can be derived directly from (\ref{eq:sshspectrum}) as $\{E_{\pm}^{\frac{1}{3}}(k),\omega_3E_{\pm}^{\frac{1}{3}}(k),\omega_3^{-1}E_{\pm}^{\frac{1}{3}}(k)\}$.
In Figs.~\ref{fig:2}(a)-(c), we represent this bulk energy spectrum for different values of $\sqrt[3]{t_1}$, after setting $\sqrt[3]{t_2}=1$.
The three branch structure, indicated by the different colors, is clearly visible.
From one branch, the other two can be obtained from successive $\frac{2\pi}{3}$ rotations in the complex energy plane, as a consequence of the $\mathscr{C}_3$-symmetry in (\ref{eq:genchiral}).
Additionally, the low energy bands of the three branches become degenerate at $E=0$ for $k=0$.
As we detail in Supplementary Section~\ref{suppl-sec:ssh3}, this actually corresponds to an exceptional point of the spectrum, with only two associated eigenstates.
The downshifted three-fold degenerate SSH real spectrum of Fig.~\ref{fig:2}(d) was obtained by cubing the complex spectrum of the $\sqrt[3]{\text{SSH}}$ model in Fig.~\ref{fig:2}(a).

% FIgure
%%%%%%%%%%%%%%%%%%%%%%%%%%%%%%%%%%%%%%%%%%%%%%%%%%%%%%%%%%%%%%%%%%%%%
%%%%%%%%%%%%%%%%%%%%%%%%%%%%%%%%%%%%%%%%%%%%%%%%%%%%%%%%%%%%%%%%%%%%%
%%%%%%%%%%%%%%%%%%%%%%%%%%%%%%%%%%%%%%%%%%%%%%%%%%%%%%%%%%%%%%%%%%%%%

Remarkably, the spectral gap for the $\sqrt[3]{\text{SSH}}$ model does not fall in the conventional categories of non-Hermitian systems, namely those of point or line gaps \cite{Kawabata2019,Bergholtz2021}, which are present if the Hamiltonian can be continuously flattened into a unitary matrix without closing the respective gap.
Here, and since the starting ($\sqrt[3]{\text{SSH}}$) model is directly related to the parent (SSH) model by a cubing procedure, the energy gap of the latter [see Fig.~\ref{fig:2}(d)], present for $t_1\neq t_2$, naturally reverts back to all three branches of the complex spectrum [see Fig.~\ref{fig:2}(a)].
This generates what we label as a \textit{ring gap} in the energy spectrum, not reducible to a point or a line gap. 
In the sequence of Figs.~\ref{fig:2}(a)-(c), we can see the ring gap closing and reopening across the critical point $\sqrt[3]{t_1}=\sqrt[3]{t_2}$.
A continuous ring gap is obtained in the $n\to\infty$ limit of the $\sqrt[n]{\text{SSH}}$ model studied below, corresponding to an infinite number of branches forming a continuum in the energy spectrum (see the energy spectrum of the $n=20$ case in Supplementary Section~\ref{suppl-sec:sshn}).

Another interesting effect occurs when a $\pi$ magnetic flux is uniformly distributed in the loops of the rhombus with one type of hopping term in each unit cell.
For example, let us consider the Peierls substitution $\sqrt[3]{t_1}\to e^{i\frac{\pi}{3}}\sqrt[3]{t_1}$ at the unit cell shown at the bottom of Fig.~\ref{fig:1}.
We label the resulting system as the $\sqrt[3]{\text{SSH}}_{\frac{\pi}{3}}$ model.
As explained in more detail in Supplementary Section~\ref{suppl-sec:ssh3}, and illustrated in Figs.~\ref{fig:2}(e)-(g), when this change is included in the Hamiltonian in (\ref{eq:hamiltssh3}), it induces a $\frac{\pi}{3}$ relative rotation in the complex plane between the three outer bands and the three inner bands, and we lose the one-to-one correspondence between an outer and an inner band that previously defined each branch.
Instead, now we distinguish between outer and inner branches, which have a relative $\frac{\pi}{3}$ phase difference.

\begin{figure*}[t]
	\begin{centering}
		\includegraphics[width=0.9\textwidth]{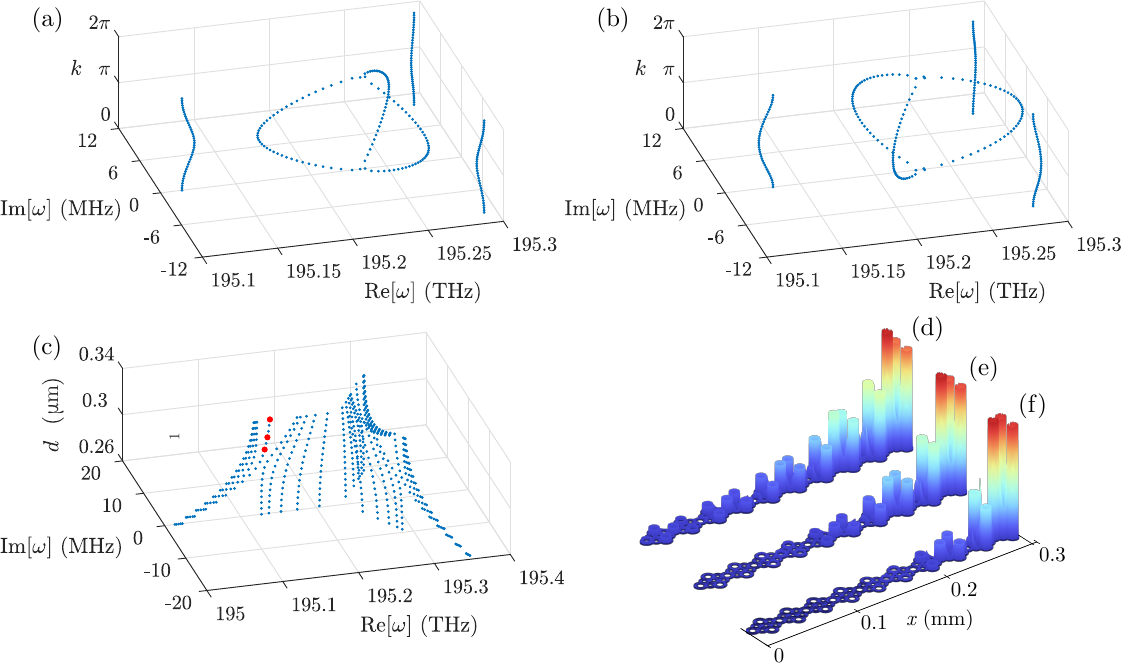} 
		\par\end{centering}
	\caption{Eigenfrequencies of the photonic (a) $\sqrt[3]{\text{SSH}}$ model and (b) $\sqrt[3]{\text{SSH}}_{\frac{\pi}{3}}$ model with PBC at steps of $\Delta k = 0.05\pi$. The three-fold splitting along the complex plane can be readily observed. 
		%(b) Eigenfrequencies when displacing the link rings corresponding, so that $\sqrt[3]{t_1} \to e^{i\frac{\pi}{3}}\sqrt[3]{t_1}$. 
		(c) Eigenspectrum of the photonic $\sqrt[3]{\text{SSH}}$ chain with OBC and $N=5$ unit cells, with $d_2 = \SI{0.3}{\micro\meter}$ and $d_1$ spanning both the topologically trivial and nontrivial phases, i.e., $d\in [\SI{0.26}{\micro\meter},\SI{0.34}{\micro\meter}]$. Three sets of edge modes appear along the ring gap. (d)--(f) Electric field norms for the edge modes marked in red in (c), for (d) $d_1 = \SI{0.315}{\micro\meter}$, (e) $d_1 = \SI{0.325}{\micro\meter}$ and (f) $d_1 = \SI{0.34}{\micro\meter}$, respectively.}
	\label{fig:3}
\end{figure*}

Notice that the cubed spectrum of Fig.~\ref{fig:2}(h) is shifted up, in relation to the one in Fig.~\ref{fig:2}(d), such that one of the three-fold degenerate bands is pushed to the positive half of the spectrum, and also that there is a relative $\pi$ sliding of the bands between the two cases.
The energy gap is open now at $k=0$ in Fig.~\ref{fig:2}(h).
Therefore, the spectral gap of the corresponding $\sqrt[3]{\text{SSH}}_{\frac{\pi}{3}}$ child model in Fig.~\ref{fig:2}(e) is also defined as a ring gap at the $k=0$ point, which is the gap closing point in Fig.~\ref{fig:2}(f), with the inner circumference of the ring reduced to a single degenerate point at zero energy.
However, the ring gap at $k=0$ gets obscured if one employs the usual projection of the whole spectrum onto the complex energy plane [e.g., the ring gaps of Fig.~\ref{fig:2}(e) and Fig.~\ref{fig:2}(g) are clearly not visible upon projecting the spectrum onto the energy plane].
This shows how the three-dimensional representation of the spectrum, as in Figs.~\ref{fig:2}(e)-(g), is required for the manifestation of the ring gap in the $\sqrt[3]{\text{SSH}}_{\frac{\pi}{3}}$ model.

\subsection{Photonic ring realization of the $\sqrt[3]{\text{SSH}}$ model}
\label{sec:photonic}

% \begin{figure*}[t]
% 	\begin{centering}
% 		\includegraphics[width=0.9\textwidth]{discrete-prov.pdf} 
% 		\par\end{centering}
% 	\caption{\dav{Provisional figure.} Discrete spectrum and edge state.}
% 	\label{fig:4}
% \end{figure*}

We consider rings of planar waveguides with a radius of $\SI{4.5}{\micro\meter}$ and a width of $\SI{250}{\nano\meter}$. For simplicity, the cores (with refractive index $\tilde{n}_{\text{core}} = \num{3}$) are surrounded by air, leading to an overall high contrast, and the considered resonant frequency is $\SI{195.225}{\tera\hertz}$. The asymmetric effective coupling is established through smaller antiresonant link rings of radius $\SI{3.24}{\micro\meter}$ and core refractive index $\tilde{n}_{\text{link}} = 3 + 0.1i\sin{\varphi}$, where $\varphi$ is the angle of the polar coordinates with origin at the center of the link ring. For these parameter values, the loss factor in $t_{\pm} = t \, e^{\pm h}$ is computed to be $h = 2.07$, implying a coupling asymmetry ratio of $\alpha \equiv t_{-}/t_{+} = 0.016$, very close to perfect unidirectionality. Proof that this ring setup generates such a coupling is provided in Supplementary Section~\ref{suppl-sec:nonreciprocity}.
We display the unit cell of the system in Fig.~\ref{fig:1}, where we have considered the counter-clockwise circulation ($m=-1$) for the coupling distribution shown in the inset. 
To achieve the staggered distribution of couplings present in the $\sqrt[3]{\text{SSH}}$ model, this structure is replicated with alternating relative ring distances $d_1 = \SI{0.33}{\micro\meter}$ and $d_2 = \SI{0.3}{\micro\meter}$ in each plaquette, which corresponds to a coupling ratio of $\sqrt[3]{t_1}/\sqrt[3]{t_2} \simeq 0.6$. These distances are drawn between the outer radii of the rings.
The method to extract the couplings and the asymmetry parameter $h$ from the spectrum of the ring resonators is described in Supplementary Sections~\ref{suppl-sec:coupling} and~\ref{suppl-sec:flux}, respectively. 
We display the bulk spectrum of eigenfrequencies for the periodic ring system in Fig.~\ref{fig:3}(a). The spectrum agrees very well with the theoretical results shown in Fig.~\ref{fig:2}(a). 
As explained above, the three-fold splitting in the complex energy plane is a consequence of the tripartite nature of the system, as it obeys the generalized chiral symmetry in (\ref{eq:genchiral}). As before, the band gap for this system can be generalized to the complex frequency spectrum context by defining a ring gap for all $|\omega|$ in a certain interval, a concept that can be directly applied to higher-root systems as well, as we will show below. On a separate note, although the spectra for both circulations in the main rings are obtained in the simulations, we only observe a doubly-degenerate joint spectrum. Reversing the circulation in the rings corresponds to a change of all coupling directions, but the cubed system in that case is still the SSH model. This necessarily implies that both circulations yield the same spectrum, as explained in Ref.~\cite{Marques2022}. The symmetries of the full $\sqrt[3]{\text{SSH}}$ system, which incorporates both $m=\pm 1$ circulations, are further detailed in Supplementary Section~\ref{suppl-sec:symmetries}.

A real flux can be added to the system by displacing the link rings orthogonally to the coupling line \cite{Hafezi2013}, which modifies the optical path in the upper and lower arms and induces a phase in the coupling, as shown in Supplemantary Section~\ref{suppl-sec:flux}. We are particularly interested in realizing the $\sqrt[3]{\text{SSH}}_{\frac{\pi}{3}}$ model by considering a $\pi$ flux around the loops involving one type of hopping terms, namely by considering the following Peierls substitution, $\sqrt[3]{t_1} \to e^{i\frac{\pi}{3}} \sqrt[3]{t_1}$. As detailed in the previous section, this implies a sign change for one of the bands of the parent SSH model. Relative to the $\sqrt[3]{\text{SSH}}$ model of Fig.~\ref{fig:3}(a), we can see in Fig.~\ref{fig:3}(b) that the flux causes a $\pi$-sliding of the outer bands and a $\pi/3$ rotation of the inner ones, again in agreement with the theoretical results of Fig.~\ref{fig:2}(e). %(Now the indexation of the bands is not obvious and it is tricky to define a clear gap between them. For this case, we may consider that the inner circumference of the ring gap reduces to $\omega \to 0$)??

As one would expect from a root TI, the existence of edge states in the $\sqrt[3]{\text{SSH}}$ model under OBC is inherited from the parent system. One of the remarkable features of the cubic-root system is that, since it possesses three times as many bands as the parent one, it will host three times as many in-gap states at one of the edges, namely the right one. 
The absence of topological states at the left edge can be understood as follows. 
Upon cubing the lattice, the resulting SSH chain at the first sublattice will have an onsite energy offset at the leftmost site, due to its lower coordination number at the cubic-root level (two missing connections at its left).
This onsite energy shift converts the left edge state into a bulk state (the converse reasoning applies to the other two pseudo-Hermitian residual chains of the cubed model, that is, it is their respective perturbations at the right edge sites that drive the formation of an in-gap state there).
This mechanism of single-edge locking of the topological modes is typical of high-root TIs, as demonstrated, e.g., for the diamond chain (a square-root model) in \cite{Marques2021}.
For a lattice of $N=5$ unit cells, keeping the relative distances in one sublattice fixed at $d_2 = \SI{0.3}{\micro\meter}$ and sweeping $d_1$ across the topological transition point yields the spectrum showcased in Fig.~\ref{fig:3}(c). The edge states are exponentially localized around one of the ends of the lattice, with the localization length growing as $d_1$ gets closer to $d_2$ and the states evolving into bulk states after crossing the critical point $d_1 = d_2$, that is, after crossing to the topologically trivial regime. Three examples for different $d_1$ are shown in Figs.~\ref{fig:3}(d)-(f), corresponding to the eigenfrequencies marked in red in Fig.~\ref{fig:3}(c). The edge states from the other two branches are also localized around the same end of the chain, albeit with different phases in the main rings.

\subsection{$\sqrt[n]{\text{SSH}}$ model} 
\label{subsec:ssh4}

\begin{figure}[b]
	\begin{centering}
		\includegraphics[width=0.45\columnwidth]{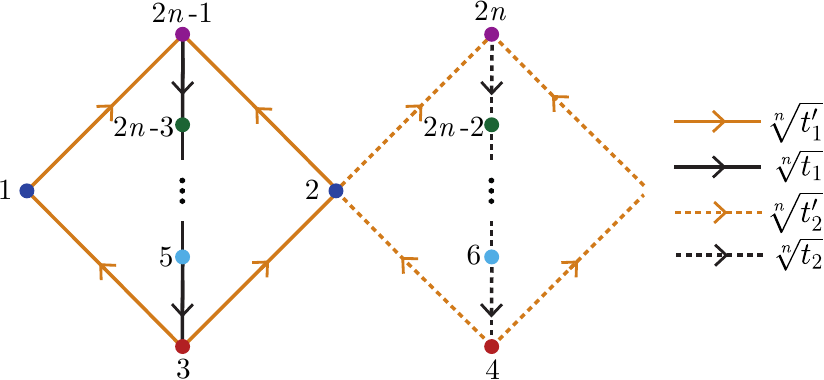} 
		\par\end{centering}
	\caption{Unit cell of the $\sqrt[n]{\text{SSH}}$ model, composed of $2n$ sites and $n$ sublattices, indicated by different colors, of two sites each.
		The arrows indicate the hopping direction, with the hopping terms assumed unidirectional. Without loss of generality, the hopping terms to or from the spinal dark blue sublattice sites can be different from the rung ones, as will be the case with the photonic ring systems studied in Section~\ref{subsec:ssh4}. As in the cubic-root case of Figs.~\ref{fig:2}(e)-(g), a $\frac{\pi}{n}$ phase shift between the two branches of $n$ bands in the complex energy spectrum can be obtained with the Peierls substitution $\sqrt[n]{t_i^{(\prime)}}\to \sqrt[n]{t_i^{(\prime)}}e^{i\frac{\pi}{n}}$, with $i=1\vee 2$.}
	\label{fig:4}
\end{figure}
\begin{figure*}[t]
	\begin{centering}
		\includegraphics[width=0.95 \textwidth]{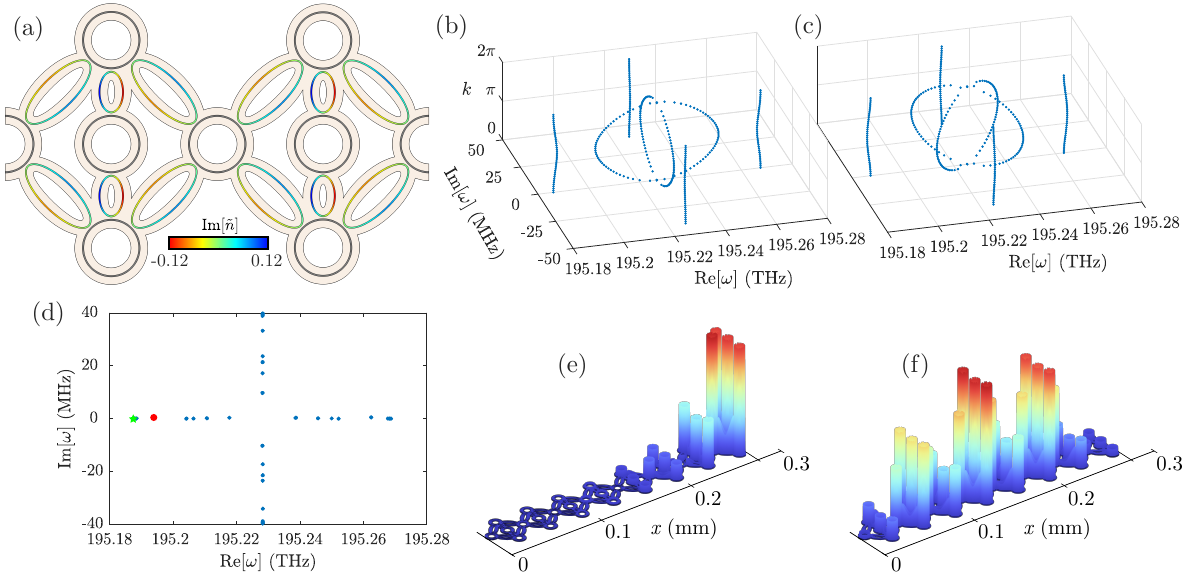} 
		\par\end{centering}
	\caption{(a) Unit cell for the $\sqrt[4]{\text{SSH}}$ model. Shorter (longer) link resonators display stronger (weaker) gain and loss modulations. 
		%(b), (c) Bulk eigenfrequencies for a periodic chain for a real flux of (b) $\phi=0$ and (c) $\phi = \pi/4$.
		Eigenfrequencies of the photonic (b) $\sqrt[4]{\text{SSH}}$ model and (c) $\sqrt[4]{\text{SSH}}_{\frac{\pi}{4}}$ model with PBC at steps of $\Delta k = 0.05\pi$.
		(d) Eigenfrequencies of the photonic $\sqrt[4]{\text{SSH}}$ chain with OBC and $N=4$ unit cells, for $d_1 = \SI{0.33}{\micro\meter}$ and $d_2 = \SI{0.3}{\micro\meter}$, where the four-fold splitting of the bands can be readily observed. (e), (f) Electric field norms for the edge and bulk modes of the system indicated by the red point and green star, respectively, in (d). }
	\label{fig:5}
\end{figure*}

As detailed in Supplementary Section~\ref{suppl-sec:sshn}, our method can be generalized to produce higher-root versions of the SSH parent model, in what we designate as the $\sqrt[n]{\text{SSH}}$ model, with integer $n>3$.
The unit cell of this system is depicted in Fig.~\ref{fig:4}.
The bulk Hamiltonian of this system, $H_{\sqrt[n]{\text{SSH}}}(k)$, exhibits $n$ two-band branches in its complex energy spectrum after diagonalization, with each branch separated from the next by a $\frac{2\pi}{n}$ angle in the energy plane due to the generalized chiral symmetry,
\begin{equation}
	\mathscr{C}_n:\ \ \Gamma_nH_{\sqrt[n]{\text{SSH}}}(k)\Gamma_n^{-1}=\omega_n^{-1}H_{\sqrt[n]{\text{SSH}}}(k),
	\label{eq:genchiralnmain}
\end{equation}
with the $\Gamma_n$ operator given in Supplementary Section~\ref{suppl-sec:sshn} and $\omega_n=e^{i\frac{2\pi}{n}}$.
After computing $H_{\sqrt[n]{\text{SSH}}}^n(k)$, 
%\dav{Maybe instead $[H_{\sqrt[n]{\text{SSH}}}(k)]^n$?} 
one obtains the Hamiltonian of the SSH model as one of its $n$ isospectral diagonal blocks.
Under OBC and for an integer number of unit cells, the $\sqrt[n]{\text{SSH}}$ model hosts $n$ edge states for the topologically non-trivial phase $t_1<t_2$, appearing at the energy gap between the two bands of each branch, which globally define the ring gap of the system. 
Finally, in the same way that as one can change from the $\sqrt[3]{\text{SSH}}$ to the $\sqrt[3]{\text{SSH}}_{\frac{\pi}{3}}$ model by introducing a $\frac{\pi}{3}$ Peierls phase at one of the hopping types (shown in Section~\ref{subsec:ssh3}), the $\sqrt[n]{\text{SSH}}_\frac{\pi}{n}$ model can be realized by applying to the hoppings in Fig.~\ref{fig:4} the transformation $\sqrt[n]{t_j^{(\prime)}}\to \sqrt[n]{t_j^{(\prime)}}e^{i\frac{\pi}{n}}$, with $j=1\vee 2$. 
Analogously, the complex energy spectrum of the $\sqrt[n]{\text{SSH}}_\frac{\pi}{n}$ model will display a global $\frac{\pi}{n}$ relative shift between the inner and outer sets of energy bands, as we show below for $n=4$ and in Supplementary Section~\ref{suppl-sec:ssh5} for $n=5$.

Implementing higher-order roots in photonic ring setups can be achieved without a significant increase in complexity by adding additional main rings to the vertical coupling link. Nonetheless, the geometrical constraints forces one to use two different elongated link rings, which are otherwise equivalent to the circular link rings in the previous section. Namely, they are antiresonant to the main rings and have a distribution of gain and loss, with maximum values that are balanced so that the nonreciprocity ratios are approximately equal in all couplings.

In the case of the $\sqrt[4]{\text{SSH}}$ model, the unit cell has the shape displayed in Fig.~\ref{fig:5}(a), where the long link rings are of elliptical shape with semiaxis lengths $R_{a1} = \SI{6.85}{\micro\meter}$ and $R_{b1} = \SI{2.5}{\micro\meter}$ and maximum loss value of $\text{Im}(\tilde{n}_{\text{link}}) = 0.072$. The short ring has semiaxis lengths $R_{a2} = \SI{3.2}{\micro\meter}$ and $R_{a2} = \SI{1.81}{\micro\meter}$ and maximum loss value of $\text{Im}(\tilde{n}_{\text{link}}) = 0.12$. We use $d_1 = \SI{0.33}{\micro\meter}$ and $d_2 = \SI{0.3}{\micro\meter}$ as alternating distances for both kinds of link rings in each plaquette. This leads to the following coupling values, using the notation indicated in Fig.~\ref{fig:4} and in units of $\sqrt[4]{t_2}$: $\sqrt[4]{t_1} = \num{0.615}$, $\sqrt[4]{t_1^\prime} = \num{0.566}$ and $\sqrt[4]{t_2^\prime} = \num{0.918}$. All these couplings have a nonreciprocity ratio of around $\alpha = 0.032$.
With these parameters, we simulate the system both under PBC and OBC. In Fig.~\ref{fig:5}(b), we show that the bulk spectrum of the photonic implementation of the $\sqrt[4]{\text{SSH}}$ model correctly captures the four-fold splitting of the bands along the complex plane, as well as the ring gap between the inner and outer bands.
A similar agreement with the theoretical result is seen in Fig.~\ref{fig:5}(c), where the bulk spectrum of the photonic $\sqrt[4]{\text{SSH}}_\frac{\pi}{4}$ is plotted. Finally, the spectrum for OBC of the photonic $\sqrt[4]{\text{SSH}}$ model with $N=4$ unit cells is shown in Fig.~\ref{fig:5}(d), where four edge modes are present, as expected. The highlighted edge mode in red is showcased in Fig.~\ref{fig:5}(e), together with a bulk mode in Fig.~\ref{fig:5}(f) for comparison.

\section{Discussion}

We have demonstrated a method to obtain general $n$-root systems of the SSH model, which requires the usage of unidirectional couplings to be implemented. This poses a challenge, as non-Hermitian systems have proven to be elusive to experimental efforts until recently, where major advances have been achieved \cite{Longhi2015,Lin2021,Lin2021OL,Zhang2021,Gu2022,Gao2022,Liu2022,Gao2023,Weidemann2020,Weidemann2022,Ke2023,Gong2018,Liu2019,Hofmann2019,Helbig2020,Liu2021,Zou2021,Qin2020,Zheng2022}. Among different possible platforms, we focused on a system of photonic ring resonators, showing it to be a very viable candidate for the implementation of $n$-root TIs, since quasi-unidirectional couplings can be realized by means of auxiliary link rings with a non-uniform imaginary component of the refractive index. Additionally, the high versatility of this platform makes it ideal for designing $n$-root systems, as it also allows, e.g., for a very precise control over the effective magnetic flux piercing the loops of these systems by simply adjusting the position of the link rings.

Implementation of systems similar to the one in this work has been accomplished with waveguide technology \cite{Hafezi2013}, where the positioning of the link rings is precise enough to allow introducing real phases in the couplings between main rings. The key challenge in our case is the correct engineering of the link rings. Non-Hermitian couplings in ring systems have already been achieved in lossy acoustic setups \cite{Zhang2021,Gao2022,Gu2022}. If no gain is considered in our system, or if gain and loss are not perfectly balanced, the effective Hamiltonian picks up imaginary diagonal elements that distort the bands. However, the main features of the model remain unaltered. We showcase this in Supplementary Section~\ref{suppl-sec:lossy}.

More recently, the split gain and loss has been implemented using optically pumped waveguides, where the lasing of different modes has been exploited \cite{Liu2022}. The effective coupling generated in that case is analogous to the one employed here, and could allow to build the root systems in an experiment. Note that the gain/loss function need not be sine-like to achieve the results in this work, although sharp transitions from gain to loss within the same ring may cause reflection effects leading to small cross-circulation couplings. This effect can cause small band splitting, but it does not distort the properties of the whole system. Note that instead one might separate the gain and loss regions into different link rings instead of within a single ring \cite{Gao2022}, or consider elongated waveguides as couplers over which the available gain can be maximized \cite{Liu2022,Gao2023}.

On the theoretical side, the method for the construction of $n$-root TIs, based on coupling loop modules of unidirectional couplings, is completely general and therefore not limited to the SSH model.
As such, our work paves the way for further studies generalizing the applicability of the method to other emblematic topological and flat-band systems, and is expected to significantly broaden the scope of high-root topology from the $2^n$-root models \cite{Dias2021,Marques2021,Marques2021b,Marques2023} studied thus far.
%Part of this work is already underway.

\section{Materials and Methods}

Numerical simulations on the ring systems are performed using the commercial finite-element simulation software COMSOL Multiphysics. All relevant parameters necessary to reproduce the results are indicated within the main text or supplementary material, or can be deduced from them. %Special care should be placed in the meshing to accurately capture the gain/loss distribution.

% SECTION
%%%%%%%%%%%%%%%%%%%%%%%%%%%%%%%%%%%%%%%%%%%%%%%%%%%%%%%%%%%%%%%%%%%%%
%%%%%%%%%%%%%%%%%%%%%%%%%%%%%%%%%%%%%%%%%%%%%%%%%%%%%%%%%%%%%%%%%%%%%
%%%%%%%%%%%%%%%%%%%%%%%%%%%%%%%%%%%%%%%%%%%%%%%%%%%%%%%%%%%%%%%%%%%%%
\section*{Acknowledgments}
\label{sec:acknowledments}

D.V. and V.A. acknowledge financial support from the Spanish State Research Agency AEI (contract No. PID2020-118153GBI00/AEI/10.13039/501100011033) and Generalitat de Catalunya (Contract No. SGR2021-00138).
A.M.M. and R.G.D. acknowledge financial support from the Portuguese Institute for Nanostructures, Nanomodelling and Nanofabrication (i3N) through Projects No. UIDB/50025/2020, No. UIDP/50025/2020, and No. LA/P/0037/2020. %, and funding from FCT-Portuguese Foundation for Science and Technology through Project
%No. PTDC/FISMAC/29291/2017. 
A.M.M. acknowledges financial support from i3N through the work Contract No. CDL-CTTRI-46-SGRH/2022.

\renewcommand{\thefigure}{S\arabic{figure}}
\renewcommand{\thetable}{S\arabic{table}}
\renewcommand{\theequation}{S\arabic{equation}}
\renewcommand{\appendixname}{Supplementary Section}
\setcounter{figure}{0} 

\begin{appendix}

\section{Cubic-root SSH} \label{suppl-sec:ssh3}

In the following, we recover and further develop the summarized analytical results on the cubic-root Su-Schrieffer-Heeger ($\sqrt[3]{\text{SSH}}$) model in Section~\ref{subsec:ssh3} of the main text. In the ordered $\{\ket{j(k)}\}$ basis, with $j=1,2,\dots,6$, the bulk Hamiltonian of the $\sqrt[3]{\text{SSH}}$ model, with the unit cell given in the lower inset of Fig.~\ref{fig:1}, can be written as
\begin{eqnarray}
	H_{\sqrt[3]{\text{SSH}}}(k)&=&
	\begin{pmatrix}
		0&h_1&0
		\\
		0&0&h_2
		\\
		h_3&0&0
	\end{pmatrix},
	\label{eq:hamiltssh3app}
	\\
	h_1&=&h_3^\dagger=-
	\begin{pmatrix}
		\sqrt[3]{t_1}&\sqrt[3]{t_2}e^{-ik}
		\\
		\sqrt[3]{t_1}&\sqrt[3]{t_2}
	\end{pmatrix},
	\label{eq:h1app}
	\\
	h_2&=&-
	\begin{pmatrix}
		\sqrt[3]{t_1}&0
		\\
		0&\sqrt[3]{t_2}
	\end{pmatrix},
	\label{eq:h2app}
\end{eqnarray}
where the lattice spacing was set to unity.
Due to its tripartite nature, defined by the sublattices $(1,2)$, $(3,4)$ and $(5,6)$ \cite{Marques2022}, this Hamiltonian obeys a generalized chiral symmetry,
\begin{eqnarray}
	\mathscr{C}_3:\ \ &\Gamma_3&H_{\sqrt[3]{\text{SSH}}}(k)\Gamma_3^{-1}=\omega_3^{-1}H_{\sqrt[3]{\text{SSH}}}(k),
	\label{eq:genchiralapp}
	\\
	&\Gamma_3&=\text{diag}(\sigma_0,\omega_3\sigma_0,\omega_3^{-1}\sigma_0),
	\label{eq:gamma3app}
\end{eqnarray}
with $\omega_3=e^{i\frac{2\pi}{3}}$ and $\sigma_0$ the $2\times2$ identity matrix.

After cubing the Hamiltonian in (\ref{eq:hamiltssh3app}) we obtain
\begin{equation}
	H_{\sqrt[3]{\text{SSH}}}^3(k)=
	\begin{pmatrix}
		H_{\text{SSH}^\prime}(k)&0&0
		\\
		0&H_2(k)&0
		\\
		0&0&H_3(k)
	\end{pmatrix},
	\label{eq:hamiltssh33app}
\end{equation}	
where
\begin{eqnarray}
	H_{\text{SSH}^\prime}(k)&=&h_1h_2h_3 \nonumber
	\\
	&=&-
	\begin{pmatrix}
		t_1+t_2&t_1+t_2e^{-ik}
		\\
		t_1+t_2e^{ik}&t_1+t_2
	\end{pmatrix}  \nonumber
	\\
	&=&-(t_1+t_2)\sigma_0+H_{\text{SSH}}(k),
 \label{eq:hamiltsshapp}
\end{eqnarray}
whose eigenstates and eigenvalues \cite{Delplace2011,Marques2020} can be readily determined as
\begin{eqnarray}
	E_{\pm}(k)&=&-t_1-t_2\pm\sqrt{t_1^2+t_2^2+2t_1t_2\cos k},
	\label{eq:sshspectrumapp}
	\\
	\ket{u_{\text{SSH}}^\pm(k)}&=&\frac{1}{\sqrt{2}}\begin{pmatrix}
		1\\
		\mp e^{i\phi(k)}
	\end{pmatrix},
	\label{eq:sshevecsapp}
	\\
	\cot \phi(k)&=&\frac{t_1}{t_2\sin k}+\cot k,
\end{eqnarray}
and
\begin{eqnarray}
	H_2(k)&=&h_2h_3h_1 \nonumber
	\\
	&=&-
	\begin{pmatrix}
		2t_1&t_1^{\frac{2}{3}}t_2^{\frac{1}{3}}(1+e^{-ik})
		\\
		t_1^{\frac{1}{3}}t_2^{\frac{2}{3}}(1+e^{ik})&2t_2
	\end{pmatrix},
\end{eqnarray}
and finally
\begin{equation}
	H_3(k)=h_3h_1h_2=H_2^\dagger(k).
	\label{eq:H3app}
\end{equation}
It can be proven \cite{Marques2022} that all diagonal blocks in (\ref{eq:hamiltssh33app}) share the same \text{real} energy spectrum given in (\ref{eq:sshspectrumapp}), such that $H_2(k)$ and $H_3(k)$ are pseudo-Hermitian matrices \cite{Mostafazadeh2002} defined as
\begin{eqnarray}
	\eta H_2(k)\eta^{-1}&=&H_2^\dagger(k)=H_3(k),
	\\
	\eta&=&\text{diag}(t_1^{-\frac{1}{3}},t_2^{-\frac{1}{3}}),
\end{eqnarray}
where (\ref{eq:H3app}) was used and $t_1,t_2>0$ is assumed.
The \textit{finite energy} eigenvectors of the $H_2(k)$ and $H_3(k)$ blocks can be expressed in terms of the $\ket{u_{\text{SSH}}^\pm(k)}$ in (\ref{eq:sshevecsapp}) \cite{Marques2022},
\begin{eqnarray}
	\ket{u_{2,R}^\pm(k)}&=&\frac{1}{\mathcal{N}_{2,R}^{\pm}(k)}h_2h_3\ket{u_{\text{SSH}}^\pm(k)},
	\label{eq:ket2app}
	\\
	\ket{u_{3,R}^\pm(k)}&=&\frac{1}{\mathcal{N}_{3,R}^{\pm}(k)}h_3\ket{u_{\text{SSH}}^\pm(k)},
	\label{eq:ket3app}
\end{eqnarray}
where $\mathcal{N}_{\mu,R}^{\pm(k)}$, with $\mu=2,3$, is a normalization constant, and $R$ stands for right eigenvector.
The non-Hermiticity of $H_\mu(k)$ implies that its left ($L$) and right (R) eigenstates, defined as
\begin{eqnarray}
	H_\mu(k)\ket{u_{\mu,R}^\pm(k)}&=&E_\pm(k)\ket{u_{\mu,R}^\pm(k)},
	\\
	H^\dagger_\mu(k)\ket{u_{\mu,L}^\pm(k)}&=&E^*_\pm(k)\ket{u_{\mu,L}^\pm(k)},
	\label{eq:lefteigapp}
\end{eqnarray}
are not the same in general.
Furthermore, one can develop the eigenvalue equation for the left eigenstate in (\ref{eq:lefteigapp}) by noticing that $H_\mu(k)=H^\dagger_{\bar{\mu}}(k)$ through (\ref{eq:H3app}), where $\bar{\mu}$ labels the Hamiltonian block other than the $\mu$-block, and that $E^*_\pm(k)=E_\pm(k)$ by the pseudo-Hermiticity condition, which imposes a real spectrum, leading to
\begin{equation}
	H_{\bar{\mu}}(k)\ket{u_{\mu,L}^\pm(k)}=E_\pm(k)\ket{u_{\mu,L}^\pm(k)},
\end{equation}
from where one can readily infer that
\begin{equation}
	\ket{u_{\mu,R}^\pm(k)}=\ket{u_{\bar{\mu},L}^\pm(k)},
	\label{eq:crossedketsapp}
\end{equation}
showing there to be a kind of ``cross-talk'' between the two pseudo-Hermitian blocks.
For each block, its complete basis is formed by the \textit{biorthogonal} basis \cite{Brody2013}, whose normalization condition for the eigenstates \cite{Kunst2018a,Bergholtz2021} reads as
\begin{equation}
	\braket{u_{\mu,L}^\sigma(k)}{u_{\mu,R}^{\sigma^\prime}(k)}=\delta_{\sigma,\sigma^\prime},\ \ \sigma,\sigma^\prime=\pm,
\end{equation}
which, through (\ref{eq:crossedketsapp}), can be rewritten as
\begin{equation}
	\braket{u_{\bar{\mu},R}^\sigma(k)}{u_{\mu,R}^{\sigma^\prime}(k)}=\delta_{\sigma,\sigma^\prime}.
	\label{eq:biorthogonalapp}
\end{equation}
By plugging (\ref{eq:ket2app})-(\ref{eq:ket3app}) into (\ref{eq:biorthogonalapp}), and using (\ref{eq:h1app})-(\ref{eq:h2app}), their respective normalization constants are found to obey the following relation,
\begin{equation}
	\mathcal{N}_{2,R}^{\pm}(k)\mathcal{N}_{3,R}^{\pm}(k)=E_\pm(k).
	\label{eq:normconstsapp}
\end{equation}
The extra condition that fixes the values of the normalization constants is derived by relating the problem back to the starting $\sqrt[3]{\text{SSH}}$ model, whose eigenstates can be grouped in three pairs of bands. 
One of these pairs has the form
\begin{equation}
	\ket{\psi_1^\pm(k)}=\frac{1}{\sqrt{3}}
 \renewcommand\arraystretch{1.6}
	\begin{pmatrix}
		\ket{u_{\text{SSH}}^\pm(k)}
		\\
		\ket{u_{2,R}^\pm(k)}
		\\
		\ket{u_{3,R}^\pm(k)}
	\end{pmatrix},
	\label{eq:ketpair1app}
\end{equation}
whose corresponding eigenvalue equation reads as
\begin{equation}
	H_{\sqrt[3]{\text{SSH}}}(k)\ket{\psi_1^\pm(k)}=E^{\frac{1}{3}}_\pm(k)\ket{\psi_1^\pm(k)},
	\label{eq:eigspair1app}
\end{equation}
where the eigenvalues are directly obtained by taking the cubic-root of the eigenvalues of the cubed system in (\ref{eq:sshspectrumapp}).
Then, by substituting (\ref{eq:ket2app})-(\ref{eq:ket3app}) in (\ref{eq:ketpair1app}) and, in turn, inserting it in (\ref{eq:eigspair1app}), one finally obtains the values of the normalization constants,
\begin{equation}
	\mathcal{N}_{2,R}^{\pm}(k)=E^{\frac{2}{3}}_\pm(k),\ \ \ \mathcal{N}_{3,R}^{\pm}(k)=E^{\frac{1}{3}}_\pm(k),
\end{equation}
which agree with (\ref{eq:normconstsapp}), allowing us to rewrite (\ref{eq:ketpair1app}) as
\begin{equation}
	\ket{\psi_1^\pm(k)}=\frac{1}{\sqrt{3}}
	\begin{pmatrix}
		\ket{u_{\text{SSH}}^\pm(k)}
		\\
		E^{-\frac{2}{3}}_\pm(k)h_2h_3\ket{u_{\text{SSH}}^\pm(k)}
		\\
		E^{-\frac{1}{3}}_\pm(k)h_3\ket{u_{\text{SSH}}^\pm(k)}
	\end{pmatrix}.
	\label{eq:ketpair1bapp}
\end{equation}
The other two pairs of eigenvalues and eigenstates can be obtained by making use of the generalized chiral symmetry of the system \cite{Marques2022}, as expressed in (\ref{eq:genchiralapp}). The eigenstates have the form
\begin{eqnarray}
	\ket{\psi_2^\pm(k)}&=&\Gamma_3\ket{\psi_1^\pm(k)}=\frac{1}{\sqrt{3}}
	\begin{pmatrix}
		\ket{u_{\text{SSH}}^\pm(k)}
		\\
		\omega_3\ket{u_{2,R}^\pm(k)}
		\\
		\omega_3^{-1}\ket{u_{3,R}^\pm(k)}
	\end{pmatrix},
\label{eq:eigspair2app}
	\\
	\ket{\psi_3^\pm(k)}&=&\Gamma^2_3\ket{\psi_1^\pm(k)}=\frac{1}{\sqrt{3}}
	\begin{pmatrix}
		\ket{u_{\text{SSH}}^\pm(k)}
		\\
		\omega_3^{-1}\ket{u_{2,R}^\pm(k)}
		\\
		\omega_3\ket{u_{3,R}^\pm(k)}
	\end{pmatrix},
\label{eq:eigspair3app}
\end{eqnarray}
whose corresponding eigenvalue equations are then written as
\begin{eqnarray}
	H_{\sqrt[3]{\text{SSH}}}(k)\ket{\psi_2^\pm(k)}&=&\omega_3E^{\frac{1}{3}}_\pm(k)\ket{\psi_2^\pm(k)},
	\label{eq:ener13omegaapp}
	\\
	H_{\sqrt[3]{\text{SSH}}}(k)\ket{\psi_3^\pm(k)}&=&\omega_3^{-1}E^{\frac{1}{3}}_\pm(k)\ket{\psi_3^\pm(k)}.
	\label{eq:ener13omegaminus1app}
\end{eqnarray}
As mentioned in Section~\ref{subsec:ssh3} of the main text, the three-fold degenerate zero-energy point at $k=0$ [see Fig.~\ref{fig:2}(a)] corresponds to an exceptional point of the spectrum \cite{Bergholtz2021}, with only two associated eigenstates.
One of them, $\ket{\psi_1^-(0)}=\frac{1}{\sqrt{2}}(1,-1,0,0,0,0)^T$, only has weight on the first sublattice, while the other, twice degenerate, $\ket{\psi_2^-(0)}=(t_1^{\frac{2}{3}}+t_2^{\frac{2}{3}})^{-\frac{1}{2}}(0,0,\sqrt[3]{t_2},-\sqrt[3]{t_1},0,0)^T$, only has weight on the second sublattice.

As also stated in Section~\ref{subsec:ssh3} of the main text, the bulk energy spectrum displays a ring energy gap for $\sqrt[3]{t_1}\neq\sqrt[3]{t_2}$, which is irreducible to the point or line gaps discussed in the literature \cite{Kawabata2019,Bergholtz2021}. The explicit formula for the ring energy gap magnitude is given by 
\begin{equation}
	\Delta=E_R-E_r,
\end{equation}
where $E_R=\sqrt[3]{2t_{\text{max}}}$ and $E_r=\sqrt[3]{2t_{\text{min}}}$ are the outer and inner energy radii of the ring gap, respectively, with $t_{\text{max}}=\max(t_1,t_2)$ and $t_{\text{min}}=\min(t_1,t_2)$.
This completes the analytical description of the bulk $\sqrt[3]{\text{SSH}}$ model.

From the form of $E_\pm(k)$ in (\ref{eq:sshspectrumapp}), it is straightforward to check, by considering the band limits occurring at the inversion-invariant momenta $k=0,\pi$, that $E_\pm(k)\in\mathbb{R}^-_0$.
In particular, we have $E_+(k)\in[-2t_{\text{min}},0]$, as exemplified in Fig.~\ref{fig:2}(d), which can be rewritten as $E_+(k)\in e^{i\pi}[0,2t_{\text{min}}]$.
Then, by taking the cubic-root of the energy of this band, one obtains
\begin{equation}
	E_+^{\frac{1}{3}}(k)\in e^{i\phi_+}[0,\sqrt[3]{2t_{\text{min}}}],\ \ \phi_+=-\frac{\pi}{3},\frac{\pi}{3},\pi,
	\label{eq:energy13app}
\end{equation}
with the same reasoning applying to the $E_-(k)$ band, that is, there is a $\phi_-=\phi_+$ phase factor appearing on the energies of the $\sqrt[3]{\text{SSH}}$ model due to the negative sign of the energies in the cubed SSH model.
Due to the $\mathscr{C}_3$-symmetry of the model, one can choose any of the allowed $\phi_\pm$ in (\ref{eq:energy13app}), which basically selects one of the three energy branches [see examples of the three-branch structure in Figs.~\ref{fig:2}(a)-(c)], for the definition of $E_\pm^{\frac{1}{3}}(k)$ in (\ref{eq:eigspair1app}).
The other energy branches are automatically obtained with (\ref{eq:ener13omegaapp})-(\ref{eq:ener13omegaminus1app}).

Let us now perform the following Peierls substitution, $\sqrt[3]{t_1}\to \text{Exp}[i\frac{\pi}{3}(1+2q)]\sqrt[3]{t_1}$, with $q\in\mathbb{Z}$, in (\ref{eq:hamiltssh3app}).
After cubing the Hamiltonian and diagonalizing the first diagonal block, relative to the parent SSH model, we get
\begin{equation}
	E_{\pm}(k)=t_1-t_2\pm\sqrt{t_1^2+t_2^2-2t_1t_2\cos k}.
	\label{eq:sshspectrumpiover3app}
\end{equation}
Comparing with (\ref{eq:sshspectrumapp}), we see that the sign of the last term of the argument of the square-root is flipped, amounting to a $k\to k+\pi$ sliding of both energy bands, and that $t_1$ in the constant energy shift coming from the identity proportional term in the Hamiltonian of (\ref{eq:hamiltsshapp}) now has a \textit{positive} sign.
As shown in Fig.~\ref{fig:2}(h) for $t_1=0.5$, the global energy shift of $2t_1$, in relation to Fig.~\ref{fig:2}(d), pushes one of the three-fold degenerate bands to the positive half of the spectrum.
As a consequence, (\ref{eq:energy13app}) gets modified as
\begin{equation}
	E_+^{\frac{1}{3}}(k)\in e^{i\phi_+}[0,\sqrt[3]{2t_{\text{min}}}],\ \ \phi_+=-\frac{2\pi}{3},0,\frac{2\pi}{3},
	\label{eq:energy13bapp}
\end{equation}
while $\phi_-=-\frac{\pi}{3},\frac{\pi}{3},\pi$, since $E_-(k)$ is still in the negative part of the energy spectrum.
Again, the choice of $\phi_+$ and $\phi_-$ within the respective allowed sets is arbitrary since, after the choice is made, all other solutions are obtained by acting with $\Gamma_3$, as shown in (\ref{eq:eigspair2app})-(\ref{eq:ener13omegaminus1app}).
However, while for the previous case of zero Peierls phases in the hopping parameters one can choose $\phi_+=\phi_-$ [which allows one to formally define the energy branches, as in Figs.~\ref{fig:2}(a)-(c)], now this is not possible since $\phi_+\neq\phi_-$, regardless of how one chooses the phases.
Therefore, one cannot define now energy branches, but only distinguish between the inner and outer sets of three energy bands that become degenerate upon cubing the spectrum, as illustrated by the color scheme in Figs.~\ref{fig:2}(e)-(g), where a $\frac{\pi}{3}$ relative phase shift can be seen between the sets, and the relevant ring gap appears now at $k=0$ for $t_1\neq t_2$.

Finally, we note that we do not consider general distributions of Peierls phases at the hopping terms outside the set defined above (\ref{eq:sshspectrumpiover3app}).
This is due to the fact that, for general distributions, one does not recover the purely real energy spectrum of an SSH model upon cubing the system, and we can no longer relate the features of the cubic-root model to a known topological and Hermitian parent model.
We leave the study of $n$-root systems with non-Hermitian parent models for future studies.

\subsection{Deviation from unidirectionality} \label{suppl-sec:unidirectionality}

\begin{figure*}[t]
	\begin{centering}
		\includegraphics[width=0.975 \textwidth,height=3.3cm]{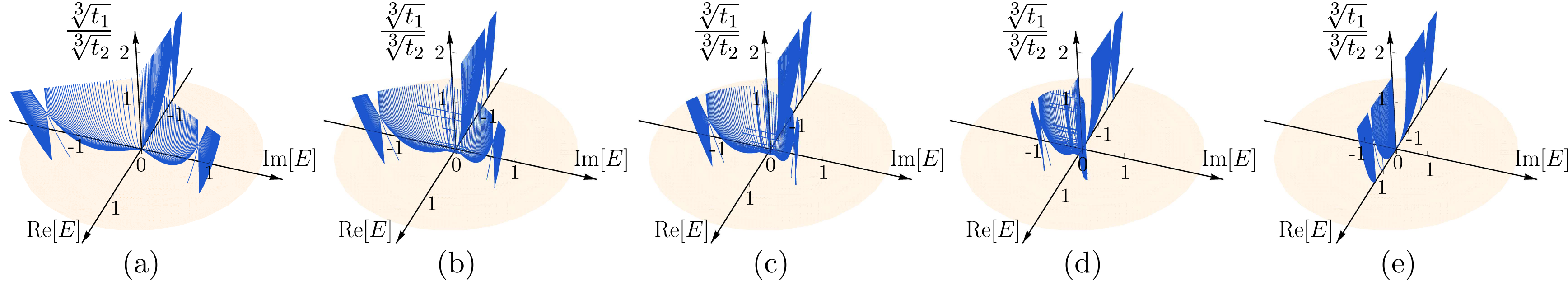} 
		\par\end{centering}
	\caption{Complex energy spectrum, in units of $\sqrt[3]{t_2}\equiv 1$, as a function of the hopping ratio obtained from diagonalization of the Hamiltonian in (\ref{eq:hamiltssh3dev}) for an open chain with $N=40$ unit cells and (a) $\alpha=0$, (b) $\alpha=0.13$, (c) $\alpha=0.25$, (d) $\alpha=0.5$, and (e) $\alpha=1$, where $\alpha\equiv t_-/t_+=e^{-2h}$ is a global measure of the non-reciprocity of the hopping terms.}
	\label{fig:alpha}
\end{figure*}

The ring system of Fig.~\ref{fig:1}, with modulated gains and losses at the auxiliary rings, generates an imaginary gauge field incorporated in the effective hoppings between adjacent main rings through the $h$ parameter as $t_\pm=te^{\pm h}$.
From the Hamiltonian of the $\sqrt[3]{\text{SSH}}$ model in (\ref{eq:hamiltssh3app}), we parameterize the deviation from perfect unidirectionality as follows,
\begin{equation}
	H_{\sqrt[3]{\text{SSH}}}^\prime(k)=H_{\sqrt[3]{\text{SSH}}}(k)+\alpha H_{\sqrt[3]{\text{SSH}}}^\dagger(k),
	\label{eq:hamiltssh3dev}
\end{equation}
where $\alpha=e^{-2h}$.
Assuming the same $\alpha$ parameterization under open boundary conditions (OBC), we show in Fig.~\ref{fig:alpha} the energy spectrum of an $\sqrt[3]{\text{SSH}}$ chain with $N=40$ unit cells as a function of the coupling ratio $\sqrt[3]{t_1}/\sqrt[3]{t_2}$, for different $\alpha$ values.
As $\alpha$ is increased, little effect is observed on the real branch of the spectrum, except for some low-energy states.
The two imaginary branches, in contrast, start bending more and more towards the real energy axis, eventually collapsing on it when Hermiticity is restored for $\alpha=1$ in Fig.~\ref{fig:alpha}(e).
From the system parameters indicated at the beginning of Section~\ref{sec:photonic} in the main text, we obtain the same imaginary gauge field of $h=2.07$ for both hopping terms $\sqrt[3]{t_1}$ and $\sqrt[3]{t_2}$, translating to $\alpha=0.016$, which is very close to unidirectionality as can be seen, e.g., by comparing the numerical results of Fig.~\ref{fig:3}(a) with the analytical ones of Fig.~\ref{fig:2}(a).
We note that even for deviations as large as $\alpha=0.25$ [see Fig.~\ref{fig:alpha}(c)], the main qualitative features survive: (i) the presence of three distinct energy branches, except at the low-energy region of the spectrum, since the states near zero energy are the first ones to hybridize; (ii) the persistence of a ring gap away from the gap closing point $\sqrt[3]{t_1}=\sqrt[3]{t_2}$, harboring three in-gap edge states for the topological phase $\sqrt[3]{t_1}<\sqrt[3]{t_2}$.

% FIgure
%%%%%%%%%%%%%%%%%%%%%%%%%%%%%%%%%%%%%%%%%%%%%%%%%%%%%%%%%%%%%%%%%%%%%
%%%%%%%%%%%%%%%%%%%%%%%%%%%%%%%%%%%%%%%%%%%%%%%%%%%%%%%%%%%%%%%%%%%%%
%%%%%%%%%%%%%%%%%%%%%%%%%%%%%%%%%%%%%%%%%%%%%%%%%%%%%%%%%%%%%%%%%%%%%

\subsection{Symmetries} \label{suppl-sec:symmetries}

Let us now address the symmetries of the \textit{full} $\sqrt[3]{\text{SSH}}$ model constructed with the photonic ring system depicted in Fig.~\ref{fig:1}. 
We recall that the main rings support states with opposite circulations $m=\pm 1$.
The Hilbert space is enlarged as $\{\ket{m,j(k)}\}=\{\ket{m}\}\otimes\{\ket{j (k)}\}$, with $\ket{j (k)}$ defined above (\ref{eq:hamiltssh3app}).
As can be understood from Fig.~\ref{fig:1}, a switch in the  circulations at the main rings, $m=-1\to m=1$, amounts to a switch in the direction over which they connect with the losses and gains of the auxiliary rings (a connection through gains between main rings in one circulation becomes a connection through losses for the other circulation, and vice-versa).
Consequently, all hopping directions are swapped when $m=-1\to m=1$, and the corresponding Hamiltonian becomes
\begin{equation}
	H_{\sqrt[3]{\text{SSH}}}^{m=-1}(k)=H_{\sqrt[3]{\text{SSH}}}(k)\to H_{\sqrt[3]{\text{SSH}}}^{m=1}(k)=H_{\sqrt[3]{\text{SSH}}}^\dagger(k),
\end{equation}
and the total Hamiltonian, written in the enlarged Hilbert space, has the following block diagonal form 
\begin{equation}
	H_{\text{tot}}=
	\begin{pmatrix}
		H_{\sqrt[3]{\text{SSH}}}(k)&O_{6}
		\\
		O_{6}&H_{\sqrt[3]{\text{SSH}}}^\dagger(k)
	\end{pmatrix},
\label{eq:htot}
\end{equation}
where $O_q$ is the zero square-matrix of size $q$.
Recall that for the parameter values used in our numerical simulations, inter-circulation couplings, that is, those that couple the $m=1$ and $m=-1$ circulations, can be safely neglected. For much smaller rings, or when using sharper transitions between regions with strong gain and loss, small off-diagonal blocks should be considered.
\begin{table*}[htbp!] 
	\centering
	\begin{tabular}{|l|c|c|c|c|c|}
		
		\hline
		Operator & Representation & Action on $H_{\text{tot}}(k)$\\
		\hline
		$y$-reflection$^\dagger$ & $R_y=\begin{pmatrix}
			\sigma_0&0&0
			\\
			0&0&\sigma_0
			\\
			0&\sigma_0&0
		\end{pmatrix}$     & n/a\\
			\hline
	Inversion & $P(k)=\begin{pmatrix}
		0&1&0&0&0&0
		\\
		1&0&0&0&0&0
		\\
		0&0&1&0&0&0
		\\
		0&0&0&e^{-ik}&0&0
		\\
		0&0&0&0&1&0
				\\
		0&0&0&0&0&e^{-ik}
	\end{pmatrix}$     & n/a\\
		\hline
		$y$-mirror    & $M_y=\tau_x\otimes R_y$     & $M_yH_{\text{\text{tot}}}(k)M_y^{-1}=H_{\text{tot}}(k)$ \\
		\hline
		Parity$^\dagger$ & $P_{\text{tot}}(k)=\tau_x\otimes P(k)$     & $P_{\text{tot}}(k)H_{\text{tot}}(k)P_{\text{tot}}(k)^{-1}=H^\dagger_{\text{tot}}(-k)$ \\
		\hline
		Flip$^\dagger$ & $F=\tau_x\otimes I_6$     & $FH_{\text{tot}}(k)F^{-1}=H_{\text{tot}}^\dagger(k)$\\
		\hline
		%		Flip-mirror    & $FM_y=\sigma_0\otimes R_y$     & $FM_yH_{tot}(k)M_y^{-1}F^{-1}=H_{tot}^\dagger(k)$ & No  \\
		%		\hline
		Time-reversal$^\dagger$   & $T=FK$     & $TH_{\text{tot}}(k)T^{-1}=H_{\text{tot}}^\dagger(-k)$\\
		\hline
		Flip-time   & $FT=(\tau_0\otimes I_6)$K     & $FTH_{\text{tot}}(k)T^{-1}F^{-1}=H_{\text{tot}}(-k)$\\
		\hline
		Flip-inversion   & $FP_{\text{tot}}(k)=\tau_0\otimes P(k)$& $FP_{\text{tot}}(k)H_{\text{tot}}(k)P_{\text{tot}}(k)^{-1}F^{-1}=H_{\text{tot}}(-k)$\\
		\hline
		Parity-time   & $P_{\text{tot}}(k)T=\big(\tau_0\otimes P(k)\big)K$& $P_{\text{tot}}(k)TH_{\text{tot}}(k)T^{-1}P_{\text{tot}}(k)^{-1}=H_{\text{tot}}(k)$\\
		\hline
		%		Flip-mirror-time   & $FM_yT=(\sigma_0\otimes R_y)$K     & $FTH_{tot}(k)T^{-1}F^{-1}=H_{tot}(-k)$ & Yes  \\
		%		\hline
		Generalized Chiral   & $\Gamma_3^{\text{tot}}=\Gamma_3\oplus\Gamma_3^\dagger$& $\Gamma_3^{\text{tot}}H_{\text{tot}}(k)\Gamma_3^{\text{tot}^{-1}}=\omega_3^{-1}H_{\text{tot}}(k)$\\
		\hline
		Generalized Particle-hole   & $S_3^{\text{tot}}=\Gamma_3^{\text{tot}}FT$& $S_3^{\text{tot}}H_{\text{tot}}(k)S_3^{\text{tot}^{-1}}=\omega_3^{-1}H_{\text{tot}}(-k)$\\
		\hline
		
		\end{tabular}
	\caption{Operators and symmetries of $H_{\text{tot}}(k)$ in (\ref{eq:htot}), with the exception of the $y$-reflection$^\dagger$ and inversion, which are operations acting only on the position space, defined as $\{\ket{j(k)}\}$, with $j=1,2,\dots,6$ labeling the corresponding site within the unit cell. $\tau_0$ ($\tau_x$) is the $2\times 2$ identity matrix ($x$ Pauli matrix) acting on the space of circulations. $\Gamma_3$ is given in (\ref{eq:gamma3app}) and $\omega_3=e^{i\frac{2\pi}{3}}$.}	
		\label{tab:syms}
\end{table*}
In Table~\ref{tab:syms}, we summarize the symmetries of $H_{\text{tot}}(k)$, indicating the matrix representation of their respective operators.
Non-Hermitian symmetries \cite{Kawabata2019} are labeled with the ``$\dagger$'' symbol, with parity$^\dagger$ symmetry the same as the pseudo-inversion symmetry defined in \cite{Schindler2023}.

The $y$-mirror operation, $M_y$, performed about the central horizontal axis of the ring system in Fig.~\ref{fig:1}, acts as a $y$-reflection$^\dagger$ operation, labeled $R_y$, on the position space, 
\begin{equation}
	R_yH_{\sqrt[3]{\text{SSH}}}(k)R_y^{-1}=H_{\sqrt[3]{\text{SSH}}}^{\dagger}(k),
\end{equation}
while simultaneously flipping the circulations in their space.
Each of these space specific operations generates a global swap of all hopping directions, such that the two considered together act as a double negative that keeps the system invariant under $M_y$, as shown in Table~\ref{tab:syms}.

The parity$^\dagger$ symmetry involves an inversion operation within the position space, defined about the vertical axis crossing sites 3 and 5 in the unit cell at the bottom of Fig.~\ref{fig:1},
\begin{equation}
	P(k)H_{\sqrt[3]{\text{SSH}}}(k)P^{-1}(k)=H_{\sqrt[3]{\text{SSH}}}(-k).
\end{equation}
Note that the inversion operator $P(k)$, given explicitly in Table~\ref{tab:syms}, is $k$-dependent, reflecting the fact that the inversion axis is shifted from the center of the unit cell.
This has been shown to lead, in Hermitian systems, to nonquantized Zak's phases for the energy bands \cite{Marques2019,Madail2019}. 
When the complete ring system in Fig.~\ref{fig:1} is considered, it is clear that the inversion operation switches the gain and loss regions of the link rings, leading to a global flip of the hopping directions, which is equivalent to a global flip of the circulations (an $x$ Pauli matrix, $\tau_x$, in the circulation space).
As a consequence, the $k$-dependent parity$^\dagger$ symmetry is obtained, whose operator $P_{\text{tot}}(k)$ and action on $H_{\text{tot}}(k)$ are given in Table~\ref{tab:syms}.

We further define in Table~\ref{tab:syms} a flip$^\dagger$ symmetry which acts non-trivially only in the circulation space, that is, the $m=\pm 1$ circulations are flipped under the action of its operator, $F=\tau_x\otimes I_6$.
This can be understood as a global reversal of the hopping directions, such that it can also be viewed, in terms of the ring system of Fig.~\ref{fig:1}, as a global swap of the gain and loss regions within each link ring. 

Some care is needed for deriving the time-reversal operator.
It is defined as an anti-unitary operator of the form $T=UK$, where $K$ is the complex conjugation operator obeying $KO=O^*K$ and $KK^{-1}=I$, while $U$ is a unitary operator that, for a general angular momentum, and up to a global phase factor, can be written as
\begin{equation}
	U=-e^{-i\pi J_y/\hbar},
\end{equation}
where $\hbar$ is the reduced Planck's constant and $J_y$ is the $y$-component of the angular momentum operator.
Assuming a bosonic pseudospin-1 system, the basis spans $\{\ket{1,1}=(1,0,0)^T,\ket{1,0}=(0,1,0)^T,\ket{1,-1}=(0,0,1)^T\}$, where the states are written in the form $\ket{l,m}$, and one has
\begin{eqnarray}
	J_y&=&\frac{i\hbar}{\sqrt{2}}
	\begin{pmatrix}
		0&-1&0
		\\
		1&0&-1
		\\
		0&1&0
	\end{pmatrix},
\label{eq:spiny}
    \\
    U&=&
 	\begin{pmatrix}
 	0&0&1
 	\\
 	0&-1&0
 	\\
 	1&0&0
    \end{pmatrix}. 
\end{eqnarray}
It is clear that $U\ket{l,m}=(-1)^{m+1}\ket{l,-m}$, \textit{i.e.}, that $U$ reverses the circulations.
In our ring system, the accessible states are the ones with opposite circulations, $m=\pm 1$.
The $m=0$ circulation does not correspond to any propagating resonant mode in the ring system and, as such, can be dropped to get, in the $\{\ket{1,1},\ket{1,-1}\}$ subspace, $U=\tau_x$.
When the entire Hilbert space is considered, one can define an operator that flips circulations as $F=\tau_x\otimes I_6$ and, from it, the time-reversal operator as $T=FK$ (see Table~\ref{tab:syms}).
Under the time-reversal operation, and apart from the usual conjugation operation, one flips the circulations.
It should also be noted that $T^2=1$, consistent with the bosonic nature of our system.
This also explains why we had to start from the $J_y$ matrix of a pseudospin-1 particle in (\ref{eq:spiny}), and then reduce it to the relevant subspace of $m=\pm 1$ circulations, rather than considering the $J_y$ matrix of a two-state pseudospin-$\frac{1}{2}$ particle, which would lead to $T^2=-1$.
Table~\ref{tab:syms} shows that the total system has the non-Hermitian time-reversal$^\dagger$ symmetry.
When it is combined with the flip operation (which duplicates the one already contained in $T=FK$), we obtain a new kind of symmetry for the full model, which we label \textit{flip-time symmetry}, whose operator is given by the product $FT$.

The generalized chiral symmetry $\mathscr{C}_3$ is also present in the full model of (\ref{eq:htot}), and its operator, $\Gamma_3^{\text{tot}}$, is given by the direct sum of the $\mathscr{C}_3$ symmetry operators of each circulation (see Table~\ref{tab:syms}), with each being the Hermitian conjugated version of the other.
Finally, we also define in Table~\ref{tab:syms} a generalized particle-hole symmetry \cite{Marques2022}, whose operator can be written as a product of three operators, $S_3^{\text{tot}}=\Gamma_3^{\text{tot}}FT$.
Conversely, note that the operator of the generalized chiral symmetry, in the full system, is not obtained in the usual way, that is, as the product of generalized particle-hole and time-reversal symmetries, $\Gamma_3^{\text{tot}}\neq S_3^{\text{tot}}T$, but rather as the product of generalized particle-hole and flip-time symmetries, $\Gamma_3^{\text{tot}}=S_3^{\text{tot}}FT$.

\section{Higher-order roots} \label{suppl-sec:sshn}

In Fig.~\ref{fig:4}, we depict the unit cell of the $n$-root SSH ($\sqrt[n]{\text{SSH}}$) model, which is composed of $2n$ sites, connected through unidirectional couplings, and $n$ two-site sublattices.
Under periodic boundary conditions (PBC), the bulk Hamiltonian of this model, in the ordered $\{\ket{j(k)}\}$ basis, with $j=1,2,\dots,2n$ (see Fig.~\ref{fig:4}), generalizes  (\ref{eq:hamiltssh3app}) as
\begin{eqnarray}
	H_{\sqrt[n]{\text{SSH}}}(k)&=&
	\begin{pmatrix}
		&h_1&
		\\
		&&h_2
		\\
		&&&\ddots
		\\
		&&&&h_{n-1}
				\\
		h_n&&&&
	\end{pmatrix},
	\label{eq:hamiltsshn}
	\\
	h_1&=&h_n^\dagger=-
	\begin{pmatrix}
		\sqrt[n]{t_1^\prime}&\sqrt[n]{t_2^\prime} e^{-ik}
		\\
		\sqrt[n]{t_1^\prime}&\sqrt[n]{t_2^\prime}
	\end{pmatrix},
	\label{eq:h1n}
	\\
	h_l&=&-
	\begin{pmatrix}
		\sqrt[n]{t_1}&0
		\\
		0&\sqrt[n]{t_2}
	\end{pmatrix},\ \ \ l=2,3,\dots,n-1,
	\label{eq:hln}
\end{eqnarray}
where the entries not shown are zeros and the lattice spacing is set to unity. 
This constitutes a particular realization of the quasi-1D $n$-partite models studied in \cite{Marques2022}, and therefore this Hamiltonian obeys a generalized chiral symmetry,
\begin{eqnarray}
	\mathscr{C}_n:\ \ &\Gamma_n&H_{\sqrt[n]{\text{SSH}}}(k)\Gamma_n^{-1}=\omega_n^{-1}H_{\sqrt[n]{\text{SSH}}}(k),
	\label{eq:genchiraln}
	\\
	&\Gamma_n&=\text{diag}(\sigma_0,\omega_n\sigma_0,\omega_n^{2}\sigma_0,\dots,\omega_n^{n-1}\sigma_0),
\end{eqnarray}
with $\omega_n=e^{i\frac{2\pi}{n}}$.
In the language introduced by some of the authors in \cite{Marques2022}, the set
\begin{equation}
	\{H_{\sqrt[n]{\text{SSH}}}(k),\omega_n^{-1} H_{\sqrt[n]{\text{SSH}}}(k),\omega_n^{-2} H_{\sqrt[n]{\text{SSH}}}(k),\dots,\omega_n H_{\sqrt[n]{\text{SSH}}}(k)\}
\end{equation}
spans the $n$ chiral colors of the same Hamiltonian, and any color can be turned into any other by applying the $\mathscr{C}_n$-symmetry operation in (\ref{eq:genchiraln}) a given amount of times.
After raising the Hamiltonian in (\ref{eq:hamiltsshn}) to the $n$th-power we obtain the following block diagonal matrix, with each block defined in a single sublattice,
\begin{equation}
		H_{\sqrt[n]{\text{SSH}}}^n(k)=\text{diag}(H_{\text{SSH}^\prime}(k),H_2(k),\dots,H_n(k)),
	\label{eq:hamiltsshnn}
\end{equation}
where the first block is an energy shifted SSH Hamiltonian,
\begin{eqnarray}
	H_{\text{SSH}^\prime}(k)&=&h_1h_2\dots h_n \nonumber
	\\
	&=&-
	\begin{pmatrix}
		w_1+w_2&w_1+w_2e^{-ik}
		\\
		w_1+w_2e^{ik}&w_1+w_2
	\end{pmatrix}  \nonumber
	\\
	&=&-(w_1+w_2)\sigma_0+H_{\text{SSH}}(k),
\end{eqnarray}
with renormalized hopping parameters, $w_i:=t_i^{\prime^{\frac{2}{n}}}t_i^{\frac{n-2}{n}}$ for $n$ odd and $w_i\to -w_i$ for $n$ even, due to the sign convention for the hopping terms in (\ref{eq:h1n})-(\ref{eq:hln}).
Its eigenvalues and eigenstates are given in (\ref{eq:sshspectrumapp})-(\ref{eq:sshevecsapp}), with $t_i\to w_i$.
The other diagonal blocks of (\ref{eq:hamiltsshnn}) are given by
\begin{equation}
	H_j(k)=h_jh_{j+1}\dots h_{n-1+j}, \ \ j=2,3,\dots,n,
\end{equation}
respecting the periodic condition $j=n+1\to j=1$.
All these blocks are isospectral to $H_{\text{SSH}^\prime}(k)$ \cite{Marques2022} and obey the relation $H_j=H_{n+2-j}^\dagger$ (the $k$ dependence is omitted henceforth), such that they are pseudo-Hermitian matrices, except when $j=\frac{n}{2}+1$ for $n$ even, which yields a Hermitian block $H_{\frac{n}{2}+1}=H_{\frac{n}{2}+1}^\dagger$.
The eigenvalue equation for the right and left eigenstates of each of these blocks is written as
\begin{eqnarray}
	H_j\ket{u_{j,R}^\pm(k)}&=&E_\pm(k)\ket{u_{j,R}^\pm(k)},
	\\
	H_j^\dagger\ket{u_{j,L}^\pm(k)}&=&H_{n+2-j}\ket{u_{j,L}^\pm(k)}=E_\pm(k)\ket{u_{j,L}^\pm(k)},
\end{eqnarray}
from where it is clear that $\ket{u_{j,L}^\pm(k)}=\ket{u_{n+2-j,R}^\pm(k)}$, and with
\begin{equation}
	\ket{u_{j,R}^\pm(k)}=E_\pm^{-\frac{n+1-j}{n}}(k)h_jh_{j+1}\dots h_n\ket{u_{\text{SSH}}^\pm(k)},
	\label{eq:usshnr}
\end{equation}
where a finite energy, $|E_\pm (k)|>0$, is assumed.

The relations above allow us to go back and determine the analytical solutions of the starting $\sqrt[n]{\text{SSH}}$ model.
Similarly to (\ref{eq:eigspair1app}), we determine all solutions from the eigenvalue equation for branch 1,
\begin{equation}
	H_{\sqrt[n]{\text{SSH}}}(k)\ket{\psi_1^\pm(k)}=E^{\frac{1}{n}}_\pm(k)\ket{\psi_1^\pm(k)},
	\label{eq:eigspair1n}
\end{equation}
with
\begin{equation}
	\ket{\psi_1^\pm(k)}=\frac{1}{\sqrt{n}}
 \renewcommand\arraystretch{1.6}
	\begin{pmatrix}
		\ket{u_{\text{SSH}}^\pm(k)}
		\\
		\ket{u_{2,R}^\pm(k)}
		\\
		\ket{u_{3,R}^\pm(k)}
		\\
		\vdots
		\\
		\ket{u_{n,R}^\pm(k)}
	\end{pmatrix},
	\label{eq:ketpair1n}
\end{equation}
which can be rewritten, through (\ref{eq:usshnr}), as
\begin{equation}
	\ket{\psi_1^\pm(k)}=\frac{1}{\sqrt{n}}
	\begin{pmatrix}
		\ket{u_{\text{SSH}}^\pm(k)}
		\\
		E_\pm^{-\frac{n-1}{n}}(k)h_2h_3\dots h_n\ket{u_{\text{SSH}}^\pm(k)}
		\\
		E_\pm^{-\frac{n-2}{n}}(k)h_3\dots h_n\ket{u_{\text{SSH}}^\pm(k)}
		\\
		\vdots
		\\
		E_\pm^{-\frac{1}{n}}(k)h_n\ket{u_{\text{SSH}}^\pm(k)}
	\end{pmatrix}.
	\label{eq:ketpair1nb}
\end{equation}
The energy can be found by generalizing (\ref{eq:energy13app}) as $E_+^{\frac{1}{n}}(k)\in e^{i\phi_+}[0,\sqrt[n]{2w_{\text{min}}}]$, with $w_{\text{min}}=\min(w_1,w_2)$ and
\begin{equation}
	\begin{cases}
	\phi_+=\frac{\pi}{n}(1+2l)\ (\text{mod}\ 2\pi),\ \ l=0,1,\dots,n-1,\ n\ \text{odd},
	\\
	\phi_+=\frac{2\pi}{n}l\ (\text{mod}\ 2\pi),\ \ l=0,1,\dots,n-1,\ n\ \text{even},
	\end{cases}
	\label{eq:phin}
\end{equation}
with $\phi_-=\phi_+$.
As before, one can choose any $\phi_\pm$ from the allowed set, that is, any $l$ in (\ref{eq:phin}), to define the complex energy phase of branch 1 in (\ref{eq:eigspair1n}) and of the respective eigenstate components in (\ref{eq:ketpair1nb}).
After setting $\phi_\pm$ as one of the possible values in (\ref{eq:phin}), the other branches are determined from successive applications of $\Gamma_n$ in (\ref{eq:genchiraln}) as
\begin{equation}
	\ket{\psi_j^\pm(k)}=\Gamma_n^{j-1}\ket{\psi_1^\pm(k)}=\frac{1}{\sqrt{n}}
	\begin{pmatrix}
		\ket{u_{\text{SSH}}^\pm(k)}
		\\
		\omega_n^{j-1}\ket{u_{2,R}^\pm(k)}
		\\
		\omega_n^{2(j-1)}\ket{u_{3,R}^\pm(k)}
		\\
		\vdots
		\\
		\omega_n^{-(j-1)}\ket{u_{n,R}^\pm(k)}
	\end{pmatrix},
	\label{eq:ketpairjn}
\end{equation}
again with $j=2,3\dots,n$, and leading to
\begin{equation}
	H_{\sqrt[n]{\text{SSH}}}(k)\ket{\psi_j^\pm(k)}=\omega_n^{j-1}E^{\frac{1}{n}}_\pm(k)\ket{\psi_j^\pm(k)}.
	\label{eq:eigspairjn}
\end{equation}
As we have seen before for the $\sqrt[3]{\text{SSH}}$ model, there will be now an $n$-fold degenerate zero-energy point at $k=0$ that corresponds to an exceptional point of the spectrum, with the same two associated eigenstates.
One of them, $\ket{\psi_1^-(0)}=\frac{1}{\sqrt{2}}(1,-1,0,0,\vec{O}_{2n-4})^T$, with $\vec{O}_{2n-4}$ a zero vector of size $2n-4$, only has weight on the first sublattice, while the other, $(n-1)$-fold degenerate, $\ket{\psi_2^-(0)}=(t_1^{\frac{2}{n}}+t_2^{\frac{2}{n}})^{-\frac{1}{2}}(0,0,\sqrt[n]{t_2},-\sqrt[n]{t_1},\vec{O}_{2n-4})^T$, only has weight on the second sublattice, since the rungs in Fig.~\ref{fig:4} effectively behave, for this eigenstate, as unidirectional Hatano-Nelson chains \cite{Hatano1996} with all the amplitude accumulated at the edge rung sites 3 and 4.

\begin{figure}[h]
	\begin{centering}
		\includegraphics[width=0.5\columnwidth]{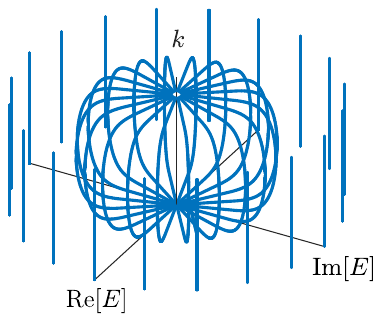} 
		\par\end{centering}
	\caption{Complex bulk energy spectrum, in  units of $\sqrt[20]{t_2} = 1$, of the $\sqrt[20]{\text{SSH}}$ model, with $\sqrt[20]{t_1} = 0.6$}
	\label{sup-fig:20root}
\end{figure}

In Fig.~\ref{sup-fig:20root}, we show the bulk energy spectrum of a very high root system, namely the $\sqrt[20]{\text{SSH}}$ model, for $\sqrt[20]{t_1} = 0.6$ and $\sqrt[20]{t_2} = 1$.
Twenty two-band branches, separated by $\frac{\pi}{10}$ angular increments in the energy plane, are present there.
The ring gap between the bottom of the outer bands and the top of the inner bands, occurring at $k=\pi$, becomes more evident as $n$ increases, becoming continuous in the $n\to\infty$ limit.

Finally, if the Peierls substitution $\sqrt[n]{t_i^{(\prime)}}\to \sqrt[n]{t_i^{(\prime)}}e^{i\frac{\pi}{n}}$ is applied to $i=1\vee 2$, then a $\pi$ flux is generated within the triangles of the plaquette with $\sqrt[n]{t_i^{(\prime)}}$ couplings in Fig.~\ref{fig:4}.
This causes a global sign shift in one of the $E_\pm(k)$ bands in $H_{\text{SSH}}^\prime$, relative to the other, which, as for the cubic-root model in Fig.~\ref{fig:2}(e)-(g), causes a $\frac{\pi}{n}$ shift between inner and outer branches of the energy spectrum.
We label this system as the $\sqrt[n]{\text{SSH}}_{\frac{\pi}{n}}$ model.
Below, we will show an example of this shift when we analyze the $n=5$ case in its photonic ring realization.
We point out that the $\sqrt[n]{\text{SSH}}$ and $\sqrt[n]{\text{SSH}}_{\frac{\pi}{n}}$ models are topologically distinct, in the sense that one cannot continuously interpolate between the two without breaking at least the $\mathscr{C}_n$-symmetry of the system.

It should also be stressed that our scheme of generating $\sqrt[n]{\text{TIs}}$ is fundamentally different from that recently proposed in \cite{Deng2022}.
There, the $n$-root of a lattice is obtained by a direct extension of the split graph \cite{Ma2020} structure of several $\sqrt{\text{TIs}}$ \cite{Ezawa2020,Mizoguchi2020,Mizoguchi2021c,Lin2021,Geng2021,Wu2021,Zhang2022,Palmer2022,Roy2022,Matsumoto2023,Geng2023}.
Namely, each link/hopping of the TI is subdivided into $n$ equal parts by adding $n-1$ sites between the two original ones.
The resulting model is always bipartite, regardless of $n$.
Upon raising its Hamiltonian to the $n$th-power, the parent TI and topologically equivalent models are obtained as diagonal blocks. However, a global off-diagonal term, labeled enhanced Hamiltonian, couples the diagonal blocks with each other in non-trivial ways, and it is not yet clear how this mixing ultimately affects the clean topological characterization of the diagonal blocks.
Our scheme for constructing $\sqrt[n]{\text{TIs}}$, on the other hand, relies on substituting the links of the parent TI with loop modules composed of unidirectional links.
The resulting system is automatically $n$-partite, according to the definition of \cite{Marques2022}, and therefore has a built-in generalized chiral symmetry $\mathscr{C}_n$.
Furthermore, when the Hamiltonian is raised to the $n$th-power, all off-diagonal blocks vanish, and a direct hierarchical relation, free of ambiguity, can be established between the $\sqrt[n]{\text{TI}}$ and the parent TI.

\subsection{Five-root model} \label{suppl-sec:ssh5}

\begin{figure*}[ht]
	\begin{centering}
		\includegraphics[width=0.95 \textwidth]{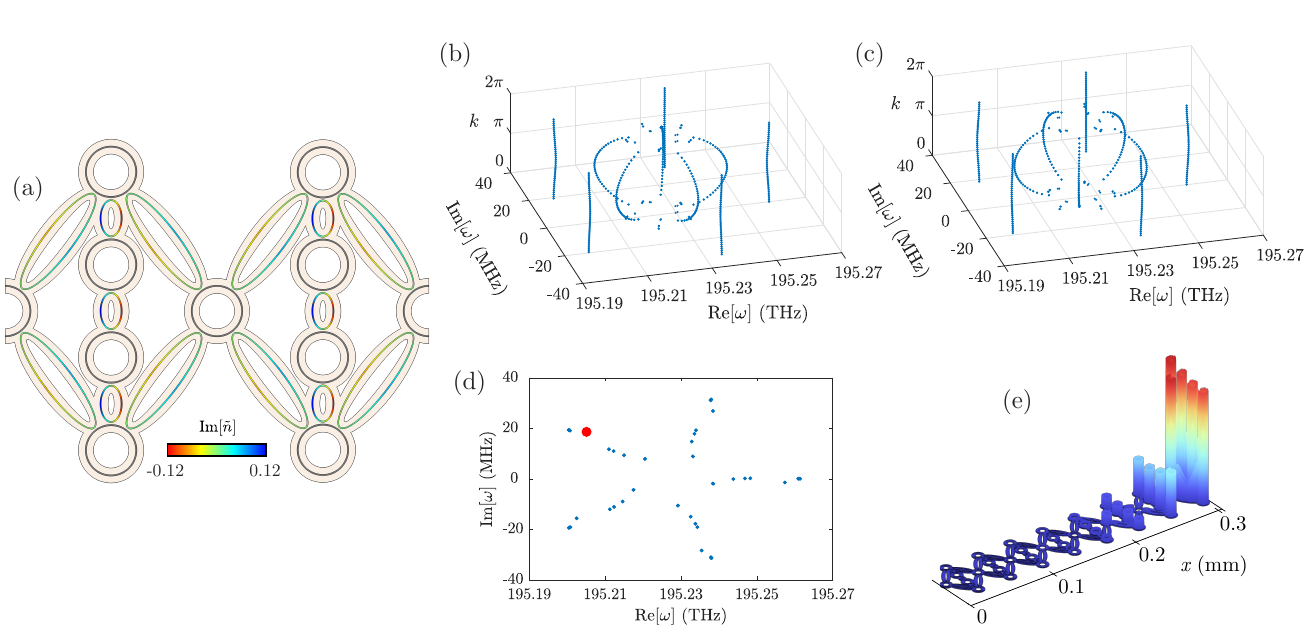} 
		\par\end{centering}
	\caption{(a) Unit cell for the $\sqrt[5]{\text{SSH}}$ model. Shorther (longer) link resonators display stronger (weaker) gain and loss values. Eigenfrequencies of the photonic (b) $\sqrt[5]{\text{SSH}}$ model and (c) $\sqrt[5]{\text{SSH}}_{\frac{\pi}{5}}$ model with PBC at steps of $\Delta k = 0.05\pi$.
		(d) Eigenfrequencies of the photonic $\sqrt[5]{\text{SSH}}$ chain with OBC and $N=4$ unit cells, for $d_1 = \SI{0.33}{\micro\meter}$ and $d_2 = \SI{0.3}{\micro\meter}$, where the five-fold splitting of the bands can be readily observed. (e) Electric field norms for the edge mode of the system indicated by the red point in (d). }
	\label{sup-fig:5r}
\end{figure*}

To demonstrate that models with a higher root degree than the ones showcased in the main text are implementable, we build the $\sqrt[5]{\text{SSH}}$ model with the unit cell displayed in Fig.~\ref{sup-fig:5r}(a). For this case, the long link ring has semiaxis lengths of $R_{a1} = \SI{10.74}{\micro\meter}$ and $R_{b1} = \SI{2.5}{\micro\meter}$ and a maximum loss value of $\text{Im}(\tilde{n}_{\text{link}}) = 0.058$, whereas the short rings are the same as the ones used in Section \ref{subsec:ssh4} in the main text. In Fig.~\ref{sup-fig:5r}(b)-(e), the same pattern of results as in the main text is repeated, now showcasing the five-fold splitting characteristic of a five-partite system. In principle, any $n$-root model can be constructed in this manner, at the cost of increasing the system size. Comparing these figures with Fig.~\ref{fig:3} and Fig.~\ref{fig:5} in the main text, corresponding to the cubic- and quartic-root models, one might note some deviations, particularly in the edge modes of Fig.~\ref{sup-fig:5r}(d) which are slightly displaced from their branches. This is mainly caused by the elongated nature of the link rings with sharp bends, which are inherently lossier than the circular rings. By virtue of the geometrical constraints, the ellipticity in this case is made large enough for distortions to appear in the spectrum. Although these are expected to increase for higher-order roots, one may move away from the elliptic design and use longer links with rounder edges to mitigate these effects.

\section{Proof of nonreciprocity} \label{suppl-sec:nonreciprocity}

\begin{figure}[h]
	\begin{centering}
		\includegraphics[width=0.7\columnwidth]{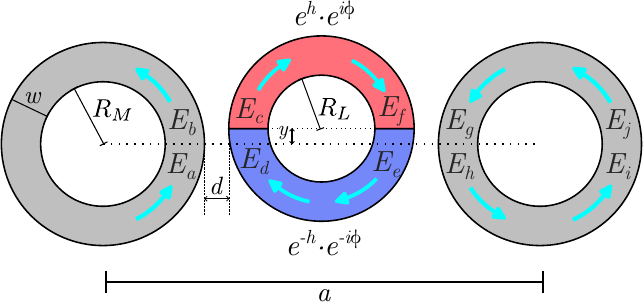} 
		\par\end{centering}
	\caption{Sketch of a set of two main rings in grey, coupled through a link ring with a split gain and loss distribution, which yields a $e^{\pm h}$ term in the coupling depending on the direction. The blue arrows indicate the circulations and the field amplitudes that are coupled in each ring. A vertical displacement from the coupling line leads to a real phase $\phi$ in the coupling.}
	\label{suppl-fig:1}
\end{figure}

As sketched in Fig.~\ref{suppl-fig:1}, we consider a set of main rings of radius $R_M$ coupled through an antiresonant link ring of radius $R_L$ with nonreciprocity parameter $h$, which is displaced a small distance $y$ from the line between the center of the two neighboring main rings so that each arm picks up a phase factor of $\phi$. All rings have a width of $w$ and we consider a separation of $d$ between their outer radii. We follow the derivation detailed in Ref.~\cite{Lin2021}, which relates the field amplitudes in neighboring rings using the transfer matrix formulation to obtain the following relations,
\begin{align}
    \begin{pmatrix}
    E_c \\ E_d \end{pmatrix}
    &= \frac{1}{i\kappa}
    \begin{pmatrix}
	\tau & -1
	\\
	1 & -\tau
    \end{pmatrix}
    \begin{pmatrix}
    E_b \\ E_a
    \end{pmatrix} = M_1\begin{pmatrix}
    E_b \\ E_a
    \end{pmatrix},
    \label{sup-eq:transfer1} \\
%\end{equation}%
%\begin{equation}
    \begin{pmatrix}
    E_f \\ E_e \end{pmatrix}
    &=
    \begin{pmatrix}
	e^{i(\beta L_L/2+\phi)}e^{h} & 0
	\\
	0 & e^{i(-\beta L_L/2+\phi)}e^{h}
    \end{pmatrix}
    \begin{pmatrix}
    E_c \\ E_d
    \end{pmatrix} = M_2 \begin{pmatrix}
    E_c \\ E_d
    \end{pmatrix},
    \label{sup-eq:transfer2} \\
%\end{equation}%
%\begin{equation}
    \begin{pmatrix}
    E_g \\ E_h \end{pmatrix}
    &= \frac{-1}{i\kappa}
    \begin{pmatrix}
	\tau & -1
	\\
	1 & -\tau
    \end{pmatrix}
    \begin{pmatrix}
    E_f \\ E_e
    \end{pmatrix} = M_3 \begin{pmatrix}
    E_f \\ E_e
    \end{pmatrix},
    \label{sup-eq:transfer3} \\
%\end{equation}%
%\begin{equation}
    \begin{pmatrix}
    E_j \\ E_i \end{pmatrix}
    &=
    \begin{pmatrix}
	e^{-i\beta L_M/2} & 0
	\\
	0 & e^{i\beta L_M/2}
    \end{pmatrix}
    \begin{pmatrix}
    E_g \\ E_h
    \end{pmatrix} = M_4\begin{pmatrix}
    E_g \\ E_h
    \end{pmatrix},
    \label{sup-eq:transfer4}
\end{align}
where $\tau$ and $\kappa$ are the transmission and coupling coefficients, $\beta$ is the propagation constant in the rings, $L_M = 2\pi[R_M+w/2]$ ($L_L$) is the circumference length of the main (link) rings and $E_l$ are the field amplitudes as labelled in Fig.~\ref{suppl-fig:1}. For a periodic system where we can apply Bloch's theorem,
\begin{equation}
    \begin{pmatrix}
    E_j \\ E_i
    \end{pmatrix} = 
    e^{ik} \begin{pmatrix}
    E_b \\ E_a
    \end{pmatrix},\label{sup-eq:bloch}
\end{equation}
where the lattice spacing is set to $a\equiv 1$. We can obtain a solution by imposing $|M_4 M_3 M_2 M_1 - e^{ik}I| = 0$, with $I$ the identity matrix. Close to resonance of the main rings, we can apply the following substitution: $\sin{(\beta L_M)} \approx (\omega-\omega_0)L_M/v_g$, where $\omega_0$ is the resonant frequency and $v_g$ the group velocity \cite{Hafezi2013,Lin2021}. Using this approximation, in addition to the antiresonant condition for the link rings, results in the following solution for the dispersion relation:
\begin{equation}
    \omega(k) = \omega_0 + t\left[e^h e^{i(\phi-k)} + e^{-h} e^{-i(\phi-k)}\right], \label{sup-eq:dispersion}
\end{equation}
where $t = v_g \kappa^2/L_M$ is the coupling strength between main rings. This dispersion relation is exactly the same as the one for a system with asymmetric coupling $t_{\pm} = t e^{\pm h} e^{\pm i \phi}$, thus proving that the considered ring setup can generate such a coupling.

\section{Extracting the coupling between rings} \label{suppl-sec:coupling}

\begin{figure}[h]
	\begin{centering}
		\includegraphics[width=0.95\columnwidth]{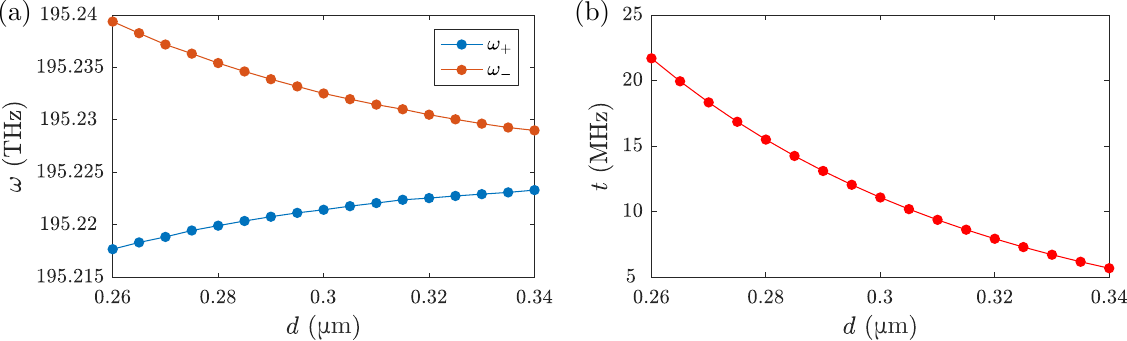} 
		\par\end{centering}
	\caption{(a) Eigenfrequencies of the set depicted in Fig.~\ref{suppl-fig:1}, composed of two main rings coupled through a link ring, and (b) coupling between main rings computed through (\ref{supp-eq:coupling2}) using the eigenfrequencies in (a), with respect to the relative distance between main and link rings.}
	\label{suppl-fig:coupling}
\end{figure}

We consider that the main rings (MRs) have a resonance frequency $\omega_0$ and that they are coupled through a link ring (LR) that creates an effective asymmetric coupling. If we focus on the basic MR-LR-MR block of Fig.~\ref{suppl-fig:1}, the Hamiltonian is given by
\begin{equation}
    H = \begin{pmatrix}
	\omega_0 & t e^{h}
	\\
	t e^{-h} & \omega_0
    \end{pmatrix},
    \label{sup-eq:coupling1}
\end{equation}
whose eigenvalues are $\omega_{\pm} = \omega_0 \pm t$. Therefore, by numerically obtaining the eigenvalues of the basic 3-ring block, we can obtain the coupling between main rings by computing
\begin{equation}
    t = \frac{1}{2}\left(\omega_{+} - \omega_{-}\right).
    \label{supp-eq:coupling2}
\end{equation}
In Fig.~\ref{suppl-fig:coupling}(a) and (b), we show the eigenvalues of the basic ring block and the coupling computed from them as a function of the distance between main and link rings, respectively.

\section{Introducing real and imaginary flux in the ring chain} \label{suppl-sec:flux}

The dispersion relation (\ref{sup-eq:dispersion}) can be expressed in the following way,
\begin{equation}
    \omega(k) = \omega_0 + 2t\cos{(k-\phi)}\cosh{h}\\ - 2it\sin{(k-\phi)}\sinh{h},\label{sup-eq:flux1}
\end{equation}
which allows us to inspect the effect of both real and imaginary flux in the system. 

\begin{figure*}[b]
	\begin{centering}
		\includegraphics[width=0.95 \textwidth]{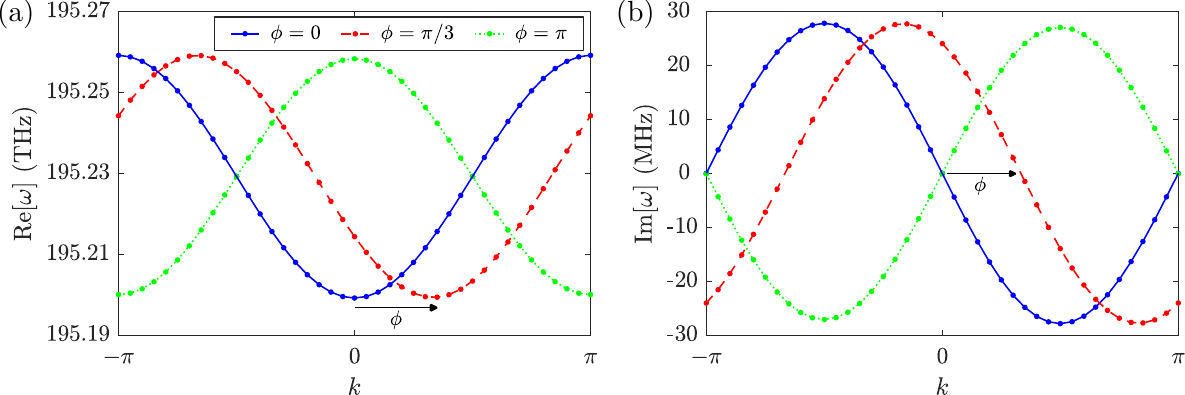} 
		\par\end{centering}
	\caption{Effect of the real flux on the real part (a) and imaginary part (b) of the eigenfrequencies of the basic MR-LR-MR block (see Fig.~\ref{suppl-fig:1}) for the counterclockwise circulation. The reverse circulation will experience the effect of an equal but opposite flux.}
	\label{sup-fig:flux}
\end{figure*}

Real flux parameterized by $\phi$ is introduced by displacing the link ring a certain distance $y$ orthogonally with respect to the line between the center of the main rings, see Fig.~\ref{suppl-fig:1}. In essence, this modifies the optical path length of the upper and lower arms of the link ring, and thus induces a phase mismatch between both geometrical paths that approximately amounts to $2\beta y$, with $\beta$ being the propagation constant of the link ring mode. This translates into a real phase $\phi$ in the coupling. From (\ref{sup-eq:flux1}), one can see that the real flux shifts both the real and imaginary parts of the spectrum along the $k$ axis, as we showcase in Fig.~\ref{sup-fig:flux}(a) and (b), respectively. Therefore, this shift allows to properly establish a relation between the flux intensity and the displacement $y$ of the link rings. Namely, for the fluxes shown in the figure, we employ a displacement of $y 
= \SI{0.169}{\micro\meter}$ ($\pi$ phase) and $y 
= 0.169/3\,\si{\micro\meter}$ ($\pi/3$ phase).

To introduce the imaginary flux, we use the split gain/loss structure explained in Section~\ref{sec:photonic} in the main text. To quantify the asymmetry with respect to this gain and loss, we set $\phi = 0$ and check that the imaginary part of (\ref{sup-eq:flux1}) has a maximum (minimum) at $k = -\pi/2$ ($k = \pi/2$). Comparing $\text{Im}(\omega)$ at these two points yields $\Delta\omega \equiv \text{Im}(\omega(-\pi/2))-\text{Im}(\omega(\pi/2)) = 4t\sinh{h}$. Therefore, if we already computed the coupling $t$ according to Supplementary Section~\ref{suppl-sec:coupling}, we can extract the parameter $h$ from
\begin{equation}
    h = \text{asinh}\left(\frac{\Delta\omega}{4t}\right). \label{sup-eq:flux2}
\end{equation}
Alternatively, by knowing that the reverse circulation experiences opposite fluxes, we can apply $h\to-h$ and $\phi\to-\phi$ to (\ref{sup-eq:flux1}). After doing this, we see that the position of the maximum and minimum for the imaginary part are reversed, so we can obtain the same result by comparing $\text{Im}(\omega)$ of both circulations  at $k=\pi/2$. From (\ref{sup-eq:flux2}), it is straightforward to compute the asymmetry ratio of couplings $\alpha \equiv t_{-}/t_{+} = e^{-2h}$ for the same circulation, which gives a notion of how close the system is to true unidirectionality.

% \begin{figure}[h]
% 	\begin{centering}
% 		\includegraphics[width=0.95 \columnwidth]{suppl-asymP.pdf} 
% 		\par\end{centering}
% 	\caption{\dav{I don't know if this one is necessary.}}
% 	\label{sup-fig:asym}
% \end{figure}

\section{Lossy ring resonators} \label{suppl-sec:lossy}

\begin{figure}[h]
	\begin{centering}
		\includegraphics[width=0.5 \columnwidth]{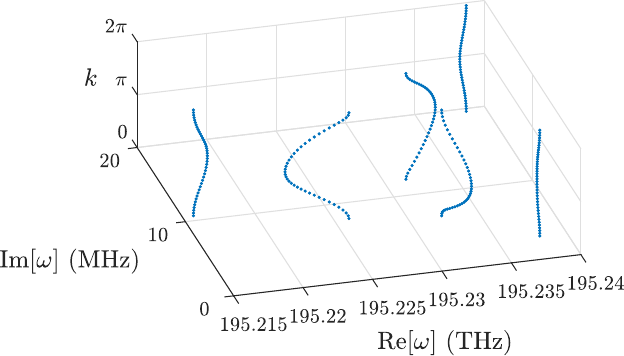} 
		\par\end{centering}
	\caption{Bulk eigenfrequencies for a periodic chain of the $\sqrt[3]{\text{SSH}}$ model using only losses in the link rings. Compared to the results in Fig.~\ref{fig:3}(a) of the main text, the bands are distorted and shifted along the complex plane, but the main qualitative features of the three-branch spectrum are still present.}
	\label{sup-fig:lossy3r}
\end{figure}

Alternatively to the split gain and loss distribution in the link rings, one may also obtain an asymmetric coupling by employing only lossy rings. The main idea is similar: the upper half of the link ring allows light to propagate and couple to the other rings, while the lower half suppresses this propagation with losses. In this situation one must note that, instead of (\ref{sup-eq:transfer2}), now that relation becomes
\begin{equation}
    \begin{pmatrix}
    E_f \\ E_e \end{pmatrix}
    =
    \begin{pmatrix}
	e^{i(\beta\pi R_2+\phi)} & 0
	\\
	0 & e^{i(-\beta\pi R_2+\phi)}e^{h}
    \end{pmatrix}
    \begin{pmatrix}
    E_c \\ E_d
    \end{pmatrix} = M_2 \begin{pmatrix}
    E_c \\ E_d
    \end{pmatrix}, \label{sup-eq:lossytransfer2}
\end{equation}
to properly describe the lack of gain in the upper path. Repeating the calculation of the dispersion relation described in Supplementary Section~\ref{suppl-sec:nonreciprocity} now yields
\begin{equation}
    \omega(k) = (\omega_0-iA) + \frac{t}{\cosh{(h/2)}}\left[e^{h/2} e^{i(\phi-k)} + e^{-h/2} e^{-i(\phi-k)}\right], \label{sup-eq:lossydispersion}
\end{equation}
where $A = 2t\tanh\left(h/2\right)$. It is implicit in (\ref{sup-eq:lossydispersion}) that, aside from the appearance of a lower coupling $t/\cosh{(h/2)}$, compared to the case with balanced gains and losses, the effective system also displays losses in the rings with a strength that is proportional to the coupling. As we see in Fig.~\ref{sup-fig:lossy3r}, this distorts the bands along the complex plane and the inner bands no longer connect at $k=0$. This can be explained by the fact that not all main rings are connected with the same number of link rings, so there are different sets of losses in the effective system. However, in the same figure we also observe that this is not enough to break the three-fold splitting of the bands, which is still maintained. For small gain/loss imbalance in the original implementation, only a small distortion of the bands is expected to occur.

\end{appendix}

\newpage

% BIBILIOGRAPHY
%%%%%%%%%%%%%%%%%%%%%%%%%%%%%%%%%%%%%%%%%%%%%%%%%%%%%%%%%%%%%%%%%%%%%
%%%%%%%%%%%%%%%%%%%%%%%%%%%%%%%%%%%%%%%%%%%%%%%%%%%%%%%%%%%%%%%%%%%%%
%%%%%%%%%%%%%%%%%%%%%%%%%%%%%%%%%%%%%%%%%%%%%%%%%%%%%%%%%%%%%%%%%%%%%

\bibliography{nrootssh}
%\cleardoublepage

\end{document}